\documentclass[structabstract]{aa}  
\usepackage{graphicx}
\usepackage{txfonts}
\usepackage{natbib}
\bibpunct{(}{)}{;}{a}{}{,} 
\begin{document}
\title{Pulsating low-mass white dwarfs in the frame of new evolutionary 
sequences} 
\subtitle{III. The pre-ELM white dwarf instability strip}
\author{A. H. C\'orsico\inst{1,2},    
        L. G. Althaus\inst{1,2},  
        A. M. Serenelli\inst{3},
        S. O. Kepler\inst{4}, 
        C. S. Jeffery\inst{5}
        \and
        M. A. Corti\inst{1,6}}
\institute{$^{1}$ Grupo  de Evoluci\'on  Estelar y  Pulsaciones,  Facultad de 
           Ciencias Astron\'omicas  y Geof\'{\i}sicas, Universidad  Nacional de
           La Plata, Paseo del Bosque s/n, (1900) La Plata, Argentina\\  
           $^{2}$ Instituto de Astrof\'{\i}sica La Plata, CONICET-UNLP, Paseo 
           del Bosque s/n, (1900) La Plata, Argentina\\
           $^{3}$ Instituto de Ciencias del Espacio (ICE-CSIC/IEEC) Campus UAB,            
           Carrer de Can Magrans, s/n 08193 Cerdanyola del Vallés, Spain\\
           $^{4}$ Departamento de Astronomia, Universidade Federal do Rio 
           Grande do Sul, Av. Bento Goncalves 9500 Porto Alegre 91501-970, 
           RS, Brazil\\
           $^{5}$ Armagh Observatory, College Hill, Armagh BT61 9DG, UK\\
           $^{6}$ Instituto  Argentino de  Radioastronom\'{\i}a, 
           CCT-La  Plata, CONICET, C.C. Nro. 5, 1984 Villa Elisa, Argentina\\
            \email{acorsico,althaus@fcaglp.unlp.edu.ar}     
           }
\date{Received ; accepted }

\abstract {Many low-mass ($M_{\star}/M_{\sun}  \lesssim 0.45$) and
  extremely low-mass (ELM, $M_{\star}/M_{\sun}  \lesssim 0.18-0.20$)
  white-dwarf stars are
  currently being found  in the  field of the Milky  Way. Some of
  these stars exhibit long-period gravity-mode ($g$-mode) pulsations, and
  constitute the class of pulsating white dwarfs called ELMV stars. In
  addition,  two low-mass pre-white dwarfs, which could be precursors
  of ELM   white dwarfs, have been observed to show  multiperiodic
  photometric variations. They could constitute a new  class of
  pulsating low-mass pre-white dwarf stars.} {Motivated by   this
  finding,  we present   a
  detailed nonadiabatic pulsation  study of such stars,
  employing full evolutionary sequences of low-mass He-core pre-white
  dwarf models.}  
  {Our pulsation stability
  analysis is based on a set of  low-mass He-core pre-white  dwarf
  models with masses  ranging  from $0.1554$   to  $0.2724  M_{\sun}$,
  which were derived by computing  the nonconservative evolution of a
  binary system consisting of an  initially $1 M_{\sun}$ ZAMS star and
  a $1.4 M_{\sun}$ neutron star companion. We have considered models in 
  which element diffusion is accounted for and also models 
  in which it is neglected.}  {We confirm and explore in detail a 
  new instability strip
  in the domain of low gravities and low effective temperatures of
  the $T_{\rm eff}-\log g$ diagram, where low-mass pre-white dwarfs are
  currently found. The destabilized modes are radial and  nonradial
  $p$ and $g$ modes excited by the  $\kappa-\gamma$
  mechanism acting mainly at the zone of the second  partial
  ionization of He, with non-negligible contributions  from the
  region of the first partial ionization  of He and the partial 
  ionization of  H. The computations with element diffusion
  are unable to explain the pulsations observed in the two known 
  pulsating pre-white 
  dwarfs, suggesting that element diffusion might be inhibited at these 
  stages of the pre-white dwarf evolution. Our nonadiabatic models 
  without diffusion, on the other hand, naturally 
  explain the existence and range   of periods of the 
  pulsating pre-white dwarf star WASP J1628$+$10B, although  
  they fail to explain the pulsations of WASP J0247$-$25B, 
  the other known member of the class,  
  indicating that  the He abundance in the driving region 
  of this star  might be substantially larger than predicted by 
  our models.}  {Further
  discoveries of additional members of this  new class  of pulsating
  stars and their analysis in the context of the theoretical
  background presented in this paper will shed new light on  the
  evolutionary history of their progenitor stars.}  
 \titlerunning{The pre-ELM white dwarf instability strip}
\authorrunning{C\'orsico et al.}
\maketitle 

%

\section{Introduction}
\label{introduction}

Most of the low- and intermediate-mass stars that populate our
Universe, including our Sun,  will become white dwarf (WD) stars  at
the very late  stages of their lives
\citep{2008ARA&A..46..157W,2008PASP..120.1043F,review}.  The majority
of WDs show H-rich atmospheres, which define the spectral class  of DA
WDs. While most of DA WDs have a  mass near  $\sim 0.60
M_{\sun}$
\citep{2007MNRAS.375.1315K,2015MNRAS.446.4078K,2011ApJ...730..128T,2013ApJS..204....5K}
and  harbor C/O cores, there is a low-mass tail of  the mass
distribution corresponding to objects with $M_{\star}/M_{\sun}
\lesssim 0.45$, which likely have cores made of He\footnote{There
  exist another component of the WD mass distribution, corresponding
  to high-mass  WDs ($M_{\star}/M_{\sun}  \gtrsim 1.0$), with cores
  probably composed of  C/O or O/Mg/Ne.}.  These low-mass WD stars are
thought to be the outcome of  strong  mass-loss events at the red
giant branch stage of  low-mass stars in binary systems  before the He
flash that, in this way, is avoided \citep{review}.  In particular,
binary evolution is the most likely origin for the  so-called
extremely low-mass  (ELM)   WDs,  which   have  masses  below   $\sim
0.18-0.20 M_{\sun}$ \citep[see the recent theoretical studies
  by][]{2013A&A...557A..19A,2014A&A...571A..45I}.    As has been
known for a long time \citep[e.g.,][]{1998A&A...339..123D}, models
predict that ELM WDs harbor very thick  H envelopes able to sustain residual H
nuclear burning via $pp-$chain, leading to markedly   long
evolutionary   timescales.  

An increasing number of low-mass  WDs, including ELM WDs,  are 
being detected through the  ELM survey  and the SPY  and  WASP
surveys   \citep{2009A&A...505..441K,  2010ApJ...723.1072B,
  2012ApJ...744..142B, 2011MNRAS.418.1156M, 2011ApJ...727....3K,
  2012ApJ...751..141K, 2013ApJ...769...66B, 2014ApJ...794...35G,
  2015MNRAS.446L..26K, 2015ApJ...812..167G}.
While these stars are already extremely
exciting  in their own right, the interest in them has been strongly
intensified because some  of them  pulsate in nonradial  $g$ (gravity)
modes \citep[ELMVs,][]{2012ApJ...750L..28H,  2013ApJ...765..102H,
  2013MNRAS.436.3573H, 2015MNRAS.446L..26K, 2015ASPC..493..217B}. This
constitutes   an unprecedented opportunity  for probing their
interiors and eventually to  test   their  formation  channels   by
employing  the   tools  of asteroseismology. Theoretical adiabatic
pulsational analysis of these stars carried out by
\citet{2010ApJ...718..441S, 2012A&A...547A..96C,  2014A&A...569A.106C}
show that $g$ modes  in ELM WDs are restricted  mainly to  the core
regions,  providing the  chance to constrain  the core  chemical
structure.   Also,  nonadiabatic  stability computations reveal that
many unstable $g$ and $p$  modes are excited by  a combination of the
$\kappa-\gamma$ mechanism  \citep{1989nos..book.....U} and the
``convective driving'' mechanism \citep{1991MNRAS.251..673B},  both of
them acting at the H-ionization  zone  \citep{2012A&A...547A..96C,
  2013ApJ...762...57V,2016A&A...585A...1C},   and   that  some
unstable    short   period    $g$    modes  can  be  driven   by    the
$\varepsilon$ mechanism due to stable H burning
\citep{2014ApJ...793L..17C}.

High-frequency pulsations in  a precursor  of a  low-mass WD star
component  of  an  eclipsing  binary  system  were  reported  for the
first time by \citet{2013Natur.498..463M}.   
This object, previously discovered by
\citet{2011MNRAS.418.1156M}, is named 1SWASP J024743.37$-$251549.2B
(hereinafter WASP J0247$-$25B). Its parameters are  
$T_{\rm  eff}=   11\,380 \pm 400$  K, $\log  g= 4.576\pm 0.011$,
and $M_{\star}=   0.186  M_{\sun}$, as given by \citet{2013Natur.498..463M}.  
It shows variability with  periods ($\Pi$) in  the range $380-420$  s,
supposed to be due  to a mixture of radial ($\ell= 0$) and nonradial 
($\ell \geq 1$) $p$ modes \citep{2013Natur.498..463M}.  Evolutionary  
models  predict that
this  star  is  evolving to  higher effective temperatures at nearly
constant luminosity prior to becoming a  low-mass WD.  This star
could be the first  member of  a new  class of pulsating  stars that
are the  precursors of  low-mass (including ELM)  WDs.  A second star
of  this   type,  1SWASP  J162842.31$+$101416.7B   (hereinafter
WASP J1628$+$10B, $T_{\rm eff}= 9\,200\pm600$ K,  $\log g=
4.49\pm0.05$,  $M_{\star}= 0.135 M_{\sun}$) was discovered by
\cite{2014MNRAS.444..208M} in another eclipsing binary  system, showing
high-frequency signals  with  periods in  the range $668-755$ s,
likely to be due    to   pulsations    similar    to   those     seen
in WASP J0247$-$25B.  The study of
\citet{2013MNRAS.435..885J}  constitutes the only theoretical work
exploring the pulsation stability  properties of radial modes of
low-mass He-core pre-WD models considering a  range  of envelope
chemical compositions,      effective       temperatures      and
luminosities.  On the basis of a huge set of static envelope models,
\citet{2013MNRAS.435..885J}  were successful to identify   the
instability  boundaries associated  with  radial  modes  characterized
by  low-to-high  radial orders, and showed that they are very
sensitive to the chemical  composition at the driving region.  These
authors   found   that   the   excitation   of   modes   is   by   the
$\kappa-\gamma$ mechanism operating mainly  in the second He
ionization zone, provided  that the  driving region  is  depleted in
H ($0.2  \lesssim X_{\rm H} \lesssim 0.3$).

In this paper,  the third work of a series devoted to low-mass WD
stars, we carry out a detailed radial and nonradial stability analysis
on a set of He-core, low-mass pre-WD evolutionary models
extracted from the computations of  \citet{2013A&A...557A..19A}. At
variance with the study of \citet{2013MNRAS.435..885J}, who employ
static envelope models, we use here fully evolutionary models that
represent the precursors  of low-mass and ELM WDs. The use of homogeneous
envelope models by \citet{2013MNRAS.435..885J}  
enables to primarily establish where pulsational
instability   might occur in the HR diagram, and allows for a full
exploration of how the pulsational instabilities  are affected  when
parameters such as the stellar mass and the  chemical composition at
the driving regions (in the stellar envelope) are varied.  
 The use of evolutionary models, on 
the other hand, involves
time-dependent evolution processes  that can have a  strong impact on
the chemical structure at the driving regions (for  instance, element
diffusion). Also, in addition to determine the instability domains and
the ranges of excited periods, the employment  of complete
evolutionary models  makes it possible to test the theories  of formation
of these stars, that is, the progenitor evolution.  Ultimately,  both
approaches ---static and evolutionary models--- are complementary. 

The paper is organized as follows. In Sect. \ref{models} we briefly
describe our numerical tools and the main ingredients of the
evolutionary sequences we employ to assess the nonadiabatic pulsation
properties of low-mass He-core pre-WDs. In Sect. \ref{stability} we
present our pulsation results in detail, devoting \S \ref{withoutdiff}
to present our non diffusion computations, and \S \ref{withdiff} to
show the results for the case in which element diffusion  is allowed
to operate. In Section \ref{observed} we compare the predictions of
our nonadiabatic models with the observed stars.  Finally, in
Sect. \ref{conclusions} we summarize the main findings of the paper.

\section{Modelling}
\label{models}

\subsection{Numerical codes}
\label{codes}

The pulsational analysis presented in this work makes use of
full stellar evolution models of pre-WDs generated  
with  the {\tt LPCODE} stellar evolution code. 
{\tt LPCODE} computes in
detail the complete evolutionary stages leading to  WD formation,
allowing one to study the WD and pre-WD evolution in a consistent way
with the expectations of the evolutionary  history of progenitors.
Details of {\tt LPCODE} can be found in
\citet{2005A&A...435..631A,2009A&A...502..207A,2013A&A...557A..19A}
and references therein. Here, we mention  only  those   ingredients
employed  which  are  relevant for  our  analysis  of  low-mass,
He-core WD and pre-WD stars \citep[see][for
  details]{2013A&A...557A..19A}.  The standard  Mixing Length  Theory
(MLT)   for convection   in the version ML2 is used
\citep{1990ApJS...72..335T}.  In this prescription, due to
\citet{1971A&A....12...21B},  the parameter $\alpha$ (the  mixing
length in units of the local pressure scale height)  is set equal to
1, while the coefficients $a, b, c$ that  appear in the equations for
the average speed of the convective cell, the average convective flux,
and  the convective efficiency \citep[see][]{1968pss..book.....C},
have values $a= 1, b= 2, c= 16$. We emphasize that the 
results presented in this work are insensitive to the prescription 
of the MLT employed. In particular, we have also used the ML1 
\citep[$\alpha= 1, a = 1/8, b = 1/2, c = 24$,][]{1958ZA.....46..108B} 
and ML3 
\citep[$\alpha= 2, a= 1, b= 2, c= 16$,][]{1990ApJS...72..335T}
recipes, and we obtain the same results than for ML2. 
The metallicity of the 
progenitor
stars has been  assumed to be $Z = 0.01$. It is worth mentioning 
that the pulsation stability results presented in this paper
do not depend on the value of $Z$\footnote{The value
of $Z$ only has a moderate impact on the value of 
the theoretical mass threshold for the development of CNO 
flashes \citep{2000MNRAS.316...84S}, which in the frame of our evolutionary 
models is $\sim 0.18 M_{\sun}$ \citep{2013A&A...557A..19A}. 
\citet{2007MNRAS.382..779P} assume $Z= 0.02$ for the progenitor 
stars and obtain a 
value of $\sim 0.17 M_{\sun}$ for the mass threshold.}. Radiative 
opacities for
arbitrary  metallicity in the range from 0 to  0.1 are from the OPAL
project \citep{1996ApJ...464..943I}.  Conductive opacities are those
of \citet{2007ApJ...661.1094C}. The  equation of state during the
main sequence evolution  is that of OPAL for H- and He-rich
compositions.  Neutrino  emission   rates   for   pair,  photo,   and
bremsstrahlung processes have been taken from
\citet{1996ApJS..102..411I},   and for plasma processes we included
the treatment of   \citet{1994ApJ...425..222H}. For the WD  regime we
have employed  an updated version of the \citet{1979A&A....72..134M}
equation of state. The nuclear network   takes into account 16
elements and 34 thermonuclear  reaction  rates   for  pp-chains,  CNO
bi-cycle,  He burning, and C ignition. Time-dependent diffusion due to
gravitational  settling and chemical  and thermal diffusion  of
nuclear  species  has been  taken into account following  the
multicomponent  gas  treatment  of
\citet{1969fecg.book.....B}. Abundance changes  have been computed
according to element diffusion, nuclear reactions,  and convective
mixing.  This detailed treatment  of abundance  changes by  different
processes during the WD regime constitutes a key aspect in the
evaluation of the importance of  residual nuclear burning for the
cooling of low-mass WDs.

We carry out a pulsation stability analysis of  radial ($\ell= 0$) and
nonradial ($\ell= 1, 2$) $p$ and $g$ modes     employing the
nonadiabatic  versions of the  {\tt LP-PUL} pulsation code described
in detail  in \citet{2006A&A...458..259C}. For the nonradial
computations, the  code solves  the  sixth-order complex system  of
linearized equations and  boundary conditions as given by
\citet{1989nos..book.....U}. For the case of radial modes,  {\tt
  LP-PUL} solves the fourth-order complex system  of linearized
equations and  boundary conditions  according to
\citet{1983ApJ...265..982S}, with the simplifications of
\citet{1993ApJ...404..294K}.  Our nonadiabatic  computations  rely on
the frozen-convection (FC) approximation,  in which  the  perturbation
of  the convective flux is neglected. While this approximation
could not be strictly valid in the case of the longest-period $g$ modes
considered,  we do not expect large variations of the nonadiabatic results 
if a time-dependent convection treatment is used
\citep[TDC;  see, for instance,][]
{2013ApJ...762...57V, 2013EPJWC..4305005S} instead, 
although we must keep in mind that this issue 
can constitute a source of uncertainties in the location of the blue edges
of instability computed in this work. 
The Brunt-V\"ais\"al\"a frequency, 
$N$,  is computed as in 
\citet{1990ApJS...72..335T}. That prescription for the computation 
of $N$ involves the so called ``Ledoux term'' $B$, which is closely 
related to the chemical gradients in the interior of the WDs and pre-WDs
\citep[see][for details]{1990ApJS...72..335T}. 

\subsection{Model sequences}
\label{model_seq}

\begin{table}
\centering
\caption{Selected  properties  of our  He-core  pre-WD  sequences: 
  the stellar  mass, the H surface abundance, the mass   
  of the H content at the point of maximum $T_{\rm eff}$
  at the beginning of the first cooling branch, 
  and the time it takes the models to evolve along the instability 
  domain (thick blue lines in Fig.\ref{figure_01}) in the non diffusion 
  computations.}
\begin{tabular}{cccc}
\hline
\hline
\noalign{\smallskip}
 $M_{\star}/M_{\sun}$  & $X_{\rm H}^{\rm surf}$ & $M_{\rm H}/M_{\sun}$ & 
$\tau_{\rm instability}$\\
& & $[10^{-3}]$ & [yr] \\ 
\noalign{\smallskip}
\hline
\noalign{\smallskip}
0.1554 & 0.365 & 4.34 &  $1.9 \times 10^{9}$ \\ 
0.1612 & 0.376 & 4.19 &  $1.3 \times 10^{9}$ \\
0.1650 & 0.390 & 4.09 &  $1.1 \times 10^{9}$ \\
0.1706 & 0.405 & 3.94 &  $4.1 \times 10^{8}$ \\
0.1762 & 0.423 & 3.80 &  $2.5 \times 10^{8}$ \\
0.1822 & 0.440 & 3.66 &  $1.6 \times 10^{8}$ \\
0.1869 & 0.453 & 3.55 &  $1.4 \times 10^{8}$ \\
0.1921 & 0.466 & 3.42 &  $5.8 \times 10^{7}$ \\
0.2026 & 0.490 & 3.22 &  $6.5 \times 10^{7}$ \\
0.2390 & 0.701 & 2.86 &  $3.4 \times 10^{7} $ \\
0.2724 & 0.715 & 2.02 &  -- \\
\hline
\hline
\end{tabular}
\label{table1}
\end{table}

\begin{figure} 
\begin{center}
\includegraphics[clip,width=9 cm]{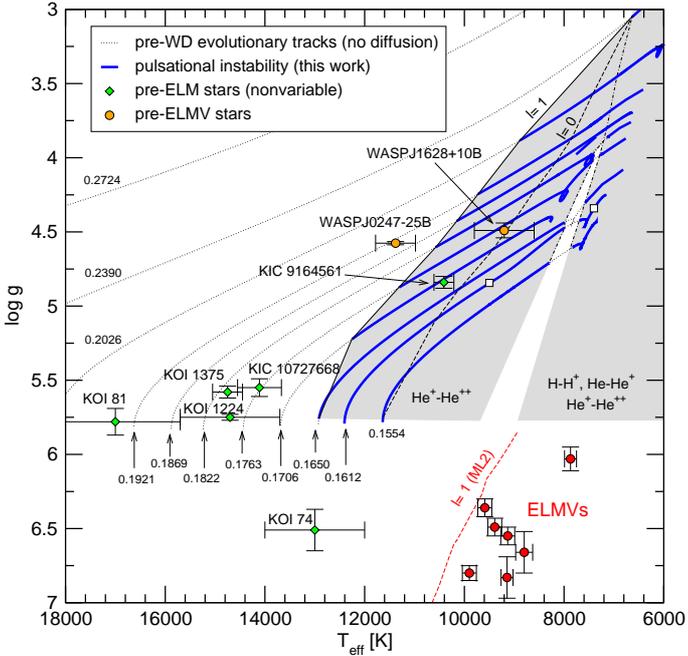} 
\caption{The $T_{\rm eff} - \log g$ diagram showing our low-mass He-core 
pre-WD evolutionary tracks (dotted curves) computed neglecting element 
 diffusion. Numbers correspond to the stellar mass of each sequence. 
The locations (with their uncertainties) of the 
  known ELMVs are marked with red dots, along with
   the theoretical $\ell= 1$ $g$-mode blue edge (red dashed curve) 
   computed by    C\'orsico \& Althaus (2015).  
   Orange dots with error bars correspond to the 
   pre-ELMV stars discovered by 
   \citet{2013Natur.498..463M,2014MNRAS.444..208M}, and
  light green diamonds depict the location of pre-ELM (nonvariable) WDs 
observed in the \emph{Kepler} mission field.
  The solid black line indicates the nonradial
  dipole ($\ell= 1$)  blue edge of the pre-ELMV  
  instability domain (emphasized as a gray area, which is an 
  extrapolation) due to 
  the $\kappa-\gamma$ mechanism acting at the He$^+-$He$^{++}$ partial 
  ionization   region, as obtained in this work. The $\ell= 2$ blue
  edge (not plotted) is nearly coincident with the $\ell= 1$ blue edge
  (see \S \ref{withoutdiff}). Similarly, the dashed black line 
  indicates the blue edge of the instability strip 
  for radial ($\ell= 0$) modes. 
  Stellar models having unstable nonradial modes are emphasized on 
  the evolutionary tracks with thick blue lines. 
  The dot-dashed black 
  lines mark the limits between the instability zones 
  in which the partial ionization of H$-$H$^+$ and He$-$He$^{+}$ 
  does contribute to the driving of modes (right zone)  and 
  does not (left zone). 
  The two  hollow squares 
  on the evolutionary track of $M_{\star}= 0.1612 M_{\sun}$ indicate the 
 location of the template models analyzed in Sect. \ref{stability}.}
\label{figure_01} 
\end{center}
\end{figure}

\citet{2013A&A...557A..19A} derived realistic  configurations   for
low-mass He-core WDs by mimicking the binary evolution  of progenitor
stars. Full details of this procedure are given  by
\citet{2013A&A...557A..19A}. 
Binary evolution was assumed to be  fully
nonconservative, and the loss of angular momentum due to  mass loss,
gravitational wave radiation, and magnetic braking was
considered. All of the He-core pre-WD initial models were derived  from
evolutionary calculations for binary systems consisting of an
evolving Main Sequence low-mass component (donor star) 
of initially $1 M_{\sun}$ and a $1.4 M_{\sun}$ neutron star companion 
as the other component. Different masses for the initial donor star 
could lead to low-mass pre-WD models with different He abundances in 
their envelopes. A total of
14 initial He-core pre-WD models with stellar masses between  $0.1554$ and
$0.4352 M_{\sun}$ were computed for initial orbital periods at the
beginning of the Roche lobe phase in the range $0.9$ to $300$ d.  In
Table \ref{table1}, we  provide some   relevant characteristics  of
the subset of  He-core  pre-WD model sequences employed in this work
which have masses in the range $0.1554 \lesssim M_{\star}/M_{\sun} 
\lesssim 0.2724$.  In 
\citet{2013A&A...557A..19A}, the complete  
evolution of these models was computed down to the range of  
luminosities of cool WDs, including the stages of multiple   
thermonuclear CNO flashes during the beginning of the cooling branch.  
In this paper, however, we focus only on the evolutionary stages 
previous to the WD evolution, that is, \emph{before} the stars reach their 
maximum effective temperature at the beginning of the first
cooling branch. The adiabatic and nonadiabatic pulsation 
properties on the final cooling branches were 
already analyzed in our previous  works of this series 
\citep{2014A&A...569A.106C,2016A&A...585A...1C}.
Column 1 of Table \ref{table1} shows
the stellar masses ($M_{\star}/M_{\sun}$). The second
column corresponds to the mass of the H content
($M_{\rm H}/M_{\star}$) at the point of maximum
effective temperature at the beginning of the first cooling branch.  
Finally,  column 4  indicates the time spent by models in 
the instability domain, marked with thick blue lines in Fig. 
\ref{figure_01}. 

\section{Stability analysis}
\label{stability}

\subsection{Computations neglecting element diffusion}
\label{withoutdiff} 

\begin{figure} 
\begin{center}
\includegraphics[clip,width=8.5 cm]{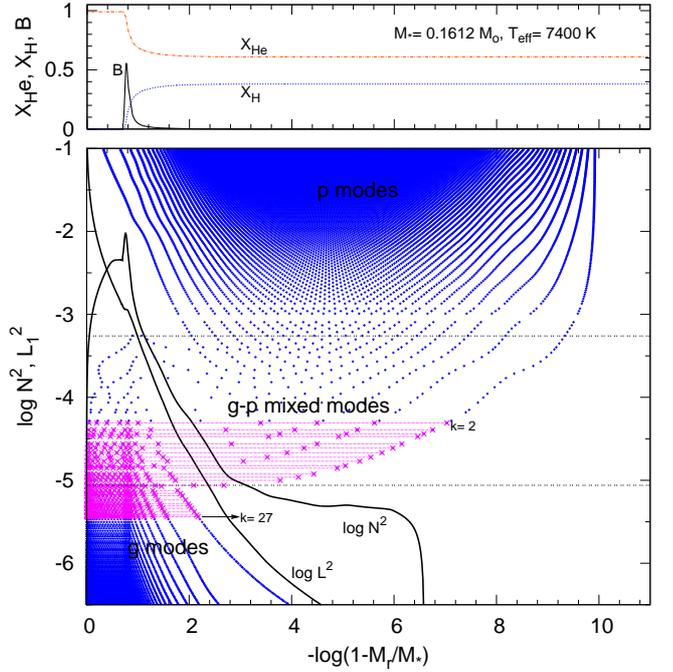} 
\caption{Internal chemical profiles of He and H and the Ledoux term $B$ 
(upper panel) and the propagation diagram --- the run of the logarithm 
of the squared critical frequencies ($N^2, L_{\ell}^2$) --- (lower panel), 
in terms of the outer mass fraction coordinate, 
corresponding to the pre-WD template model of 
$M_{\star}= 0.1612 M_{\sun}$ and $T_{\rm eff} \sim 7400$ K 
marked in Fig. \ref{figure_01}. 
In the lower panel, small ``$+$''symbols (in blue) correspond to 
the spatial location of the nodes 
of the radial eigenfunction of dipole ($\ell= 1$) $g$, $p$ and $g-p$ mixed 
modes. The (squared) frequency interval corresponding 
to $g-p$ mixed modes is enclosed with two dotted 
horizontal lines. Nodes corresponding to unstable  modes
(from the $k= 2$ $g$ mixed mode to the $k= 27$ pure $g$ mode) are 
emphasized with  ``$\times$'' symbols (in magenta) connected with thin 
lines.}
\label{figure_02} 
\end{center}
\end{figure}

\begin{figure} 
\begin{center}
\includegraphics[clip,width=8.5 cm]{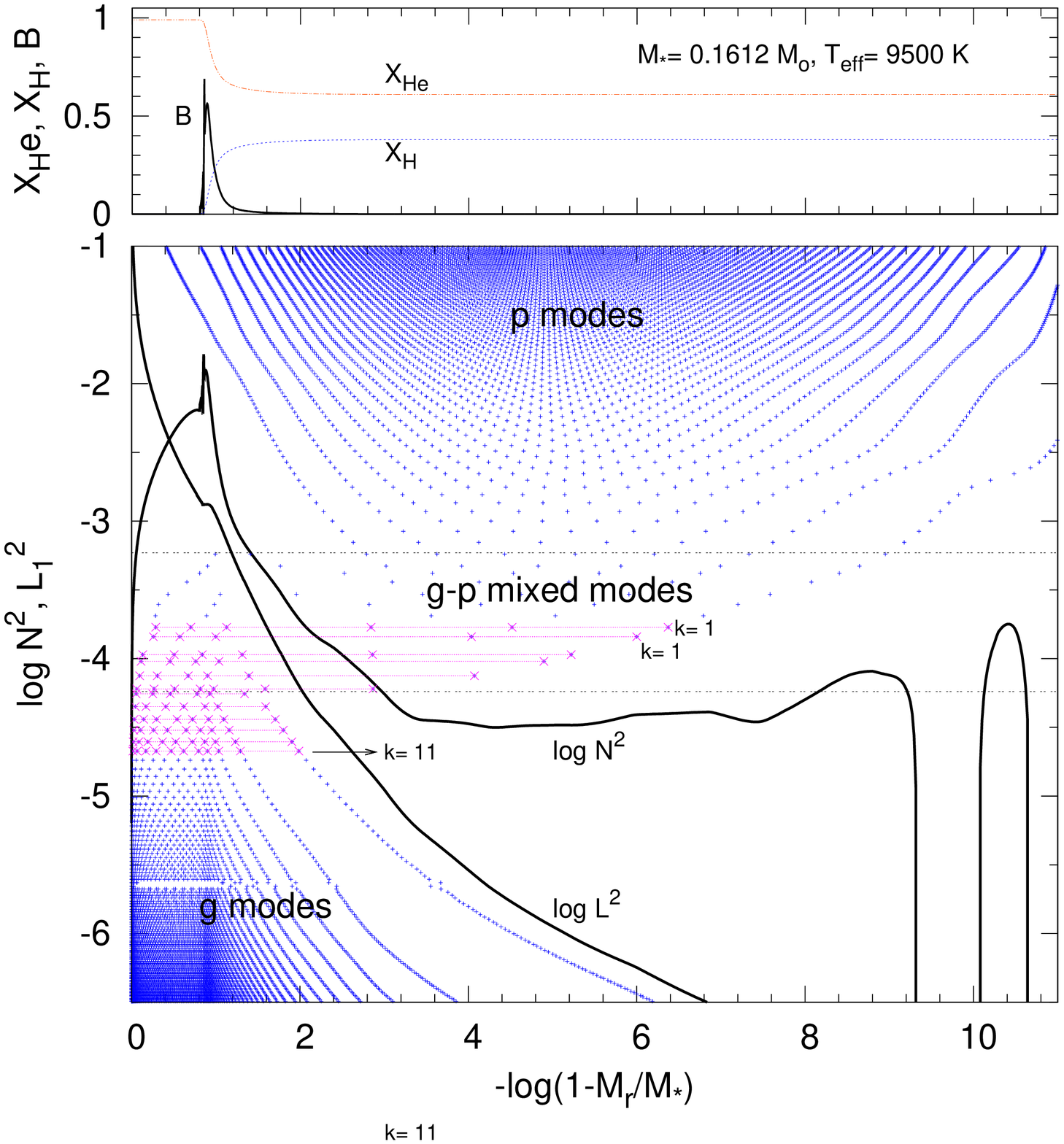} 
\caption{Same as Fig. \ref{figure_02}, but for the pre-WD template 
model of $M_{\star}= 0.1612 M_{\sun}$ and $T_{\rm eff} \sim 9500$ K 
marked in Fig. \ref{figure_01}.}
\label{figure_03} 
\end{center}
\end{figure}

We analyzed the stability pulsation properties of He-core, 
low-mass pre-WD models computed assuming the  
ML2 prescription for the MLT theory of convection
\citep[see][]{1990ApJS...72..335T} and covering a range of 
effective temperatures of $25\,000\ {\rm K} \gtrsim T_{\rm eff} 
\gtrsim 6000$ K and a range of stellar masses of 
$0.1554 \lesssim M_{\star}/M_{\sun} \lesssim 0.2724$. In this section 
we report on the results of computations in which we assume that 
element diffusion due to gravitational  settling and chemical  
and thermal diffusion is not operative. That is, it is assumed that 
each model preserves the outer He/H homogeneous chemical structure 
resulting from the previous evolution. 
We  defer the case in which time-dependent element diffusion is allowed to 
operate to \S \ref{withdiff}.
For each model, we assessed the pulsational stability of radial 
$(\ell= 0)$, and nonradial $(\ell= 1, 2)$ $p$ and $g$ modes with periods 
in the range $10\ {\rm s} \lesssim \Pi \lesssim 20\,000$ s.  
The reason for considering such a wide range of periods in our 
computations is to clearly define the theoretical domain of 
instability of pre-ELMV WDs\footnote{For simplicity, here and throughout
the paper we refer to the pulsating low-mass pre-WDs as pre-ELMV WDs,
even in the cases in which $M_{\star} \gtrsim 0.18-0.20 M_{\sun}$.}, 
that is, to find the long- and short-period edges of the instability 
domains for all the stellar masses and effective temperatures.

\begin{figure} 
\begin{center}
\includegraphics[clip,width=8.5 cm]{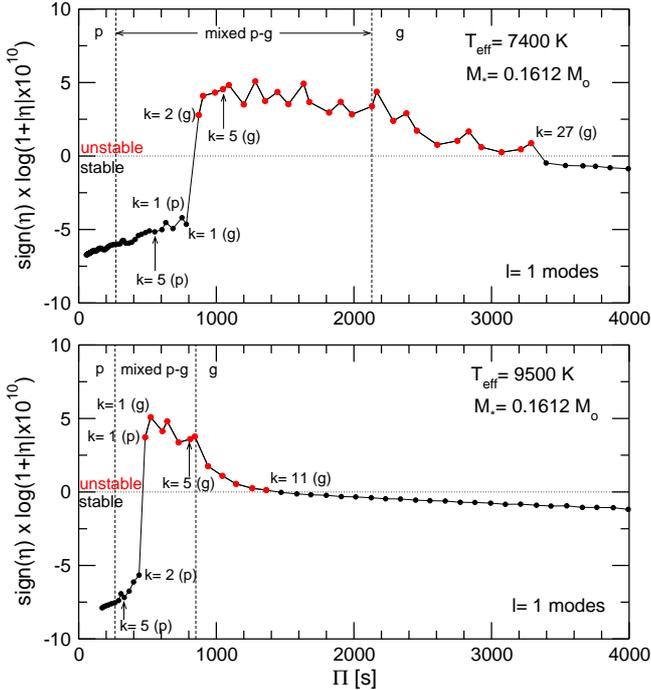} 
\caption{Normalized growth rates $\eta$ for $\ell= 1$  $p$ and $g$ modes 
in terms of the pulsation periods for the $0.1612 M_{\sun}$ pre-WD template 
model at $T_{\rm eff}= 7400$ K (upper panel) and the $0.1612 M_{\sun}$ 
pre-WD template model at $T_{\rm eff}= 9500$ K 
(lower panel). The large numerical range spanned by $\eta$ is appropriately
scaled for a better graphical representation. Some specific modes 
are labeled. The $g$ and $p$ modes with $k= 5$ are analyzed in 
Fig. \ref{figure_05}. The vertical dashed lines enclose 
the period interval of mixed $p-g$ modes. Unstable modes are emphasized 
with red dots.}
\label{figure_04} 
\end{center}
\end{figure}

We start by examining Fig. \ref{figure_01}, in which we show   the
location of the $\ell= 1$ blue (hot) edge of the instability  domain
of our low-mass He-core pre-WD models on the  $T_{\rm eff}- \log g$
plane with a solid black line. The blue edge for $\ell= 2$ modes (not 
shown in the plot) is  a little ($\sim 10-30$ K) 
hotter than the $\ell= 1$ blue edge. We emphasize that 
the location of the blue edges does not depend on the prescription
for the MLT theory of convection adopted in the equlibrium models.
The blue  edge of the instability
domain associated with radial ($\ell= 0$) modes is depicted with a
dashed black line. Visibly, the blue edge for radial modes  is
substantially cooler than for nonradial modes.  The parts of the
non-diffusion evolutionary tracks  (drawn with dotted thin lines)
associated with stellar models  having at least one unstable  nonradial
mode are emphasized with  thick blue lines. There is a strong
dependence of the blue  edges on the stellar mass, i.e., the blue
edges are hotter for the less massive model sequences. This mass
dependence is  connected with the abundance of He at the driving
regions  (see later) of the models. We also include in the plot the
blue  edge of the theoretical $\ell= 1$ ELM WD instability strip (red
dashed line)  according to C\'orsico \& Athaus (2015), along with the
seven ELMV discovered  so far \citep[][]
{2012ApJ...750L..28H,2013ApJ...765..102H,2013MNRAS.436.3573H,
2015MNRAS.446L..26K,2015ASPC..493..217B}, marked with red dots. In
comparison, the slope   of the $\ell= 1$  blue edge of the
pre-ELMVs is slightly less steep  but much hotter than the blue edge
of the ELMVs. As a result,  the instability domain of pre-ELMVs is
substantially more extended in  effective temperature. We have
also included in Fig. \ref{figure_01} the location of two pulsating pre-WD
stars discovered  by \citet{2013Natur.498..463M,2014MNRAS.444..208M}
(WASP J0247$-$25B and WASP J1628+10B, respectively).
Finally, we plot the location of six nonvariable low-mass 
pre-WDs in binary systems with companion
A stars, observed in the \emph{Kepler} field. 
They are KIC 9164561 and KIC 10727668 \citep{2015ApJ...803...82R}, 
KOI 74 and KOI 81 \citep{2010ApJ...715...51V}, 
KOI 1375 \citep{2011ApJ...728..139C}, 
and KOI 1224 \citep{2012ApJ...748..115B}.
Our current computations without element diffusion seems to  
account only for the pulsations of WASP J1628+10B, but not 
for WASP J0247$-$25B. On the other hand, our analysis 
predicts pulsational instabilities at the effective 
temperature and gravity of the constant (nonvariable)
star KIC 9164561. We will return 
to this issue in \S \ref{observed}.

\subsubsection{Propagation diagrams}

Here, we describe the nonradial stability properties of two template $0.1612
M_{\sun}$ low-mass He-core pre-WD models with $T_{\rm eff}= 7400$ K and 
$T_{\rm eff}= 9500$ K\footnote{The stability properties of radial modes
are qualitatively similar to those of nonradial modes, and therefore
they will not be described.}. Their locations in the $T_{\rm eff} - \log g$ 
diagram are displayed in Fig. \ref{figure_01} as two hollow squares
in the $0.1612 M_{\sun}$ evolutionary track. 
These  properties are representative of all the models of our 
complete set of evolutionary sequences. In the upper panel  
of Fig. \ref{figure_02} we display the
internal chemical profiles for He and H corresponding to the template
model with $T_{\rm eff}\sim 7400$ K. The model is characterized 
by a core made of pure He surrounded by an envelope with  
uniform chemical composition of H ($X_{\rm H}= 0.38$) and He 
($X_{\rm He}= 0.61$) up to the surface. In addition, we show in the 
same panel the Ledoux term $B$ \citep[related to the computation of the 
Brunt-V\"ais\"al\"a frequency; see][for its definition]
{1990ApJS...72..335T}, which  has nonzero
values only at the chemical gradient associated with the He/H  transition
region. 

\begin{figure*} 
\begin{center}
\includegraphics[clip,width=17 cm]{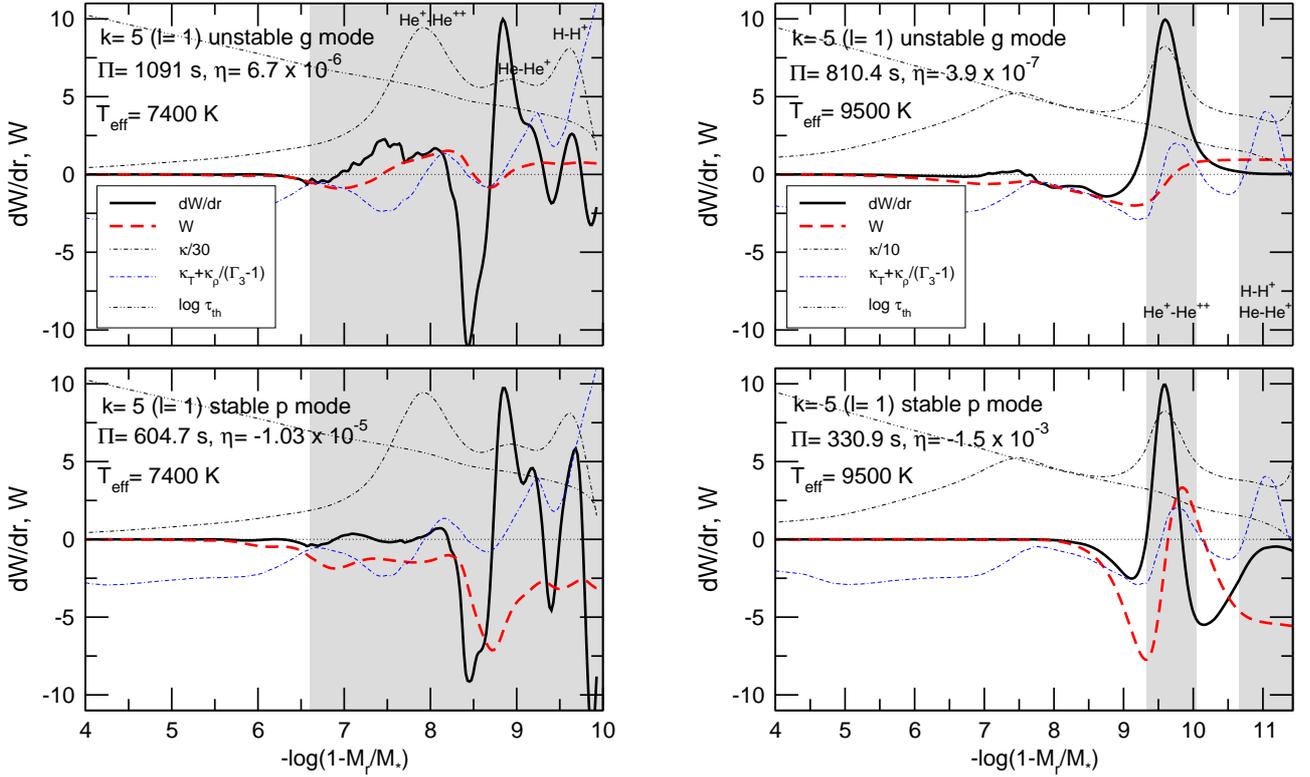} 
\caption{Left panels: the differential work ($dW/dr$) and the running 
work integral ($W$) for the unstable $g$ mode with $k= 5$ 
(upper panel) and the stable $p$ mode with $k= 5$
(lower panel), along with the Rosseland opacity profile ($\kappa$), 
the opacity derivatives, and the thermal timescale 
($\tau_{\rm th}$) of our $0.1612 M_{\sun}$ pre-WD
template model at $T_{\rm eff}= 7400$ K. The gray areas show the location of
the convection zones due to the partial ionization zones. Right panels: 
same as in the left panels, but for the $0.1612 M_{\sun}$ pre-WD
template model at $T_{\rm eff}= 9500$ K.}
\label{figure_05} 
\end{center}
\end{figure*}

The He/H chemical transition region leaves  notorious signatures in
the run of the squared critical frequencies,  in particular in the
Brunt-V\"ais\"al\"a frequency.  This is clearly displayed in the
propagation diagram \citep{1980tsp..book.....C,1989nos..book.....U} of
the lower  panel in Fig. \ref{figure_02}. $g$ modes propagate in the
regions of the star where $\sigma^2 < N^2, L^2_{\ell}$,  and $p$ modes
in the regions where $\sigma^2 > N^2, L^2_{\ell}$.  Here, $L_{\ell}$
is the Lamb (acoustic) frequency, defined as $L^2_{\ell}=
\ell(\ell+1)c_{\rm s}^2/r^2$ ($c_{\rm s}$ being the local adiabatic
sound speed), and  $\sigma$ is the oscillation frequency.  Because 
of the
very peculiar shape  of the Brunt-V\"ais\"al\"a frequency in the inner
regions of the star,  there is a considerable range of intermediate
frequencies for which the modes behave like $g$ modes in the inner
parts of the star and  like $p$ modes in the outer parts. A similar
situation is found in hydrogen-deficient pre-WD models representative
of  PG 1159 stars before the evolutionary knee  \citep[see Fig. 6
  of][]{2006A&A...454..863C}. These  intermediate-frequency modes  are
called ``mixed modes'' \citep{1974A&A....36..107S,1975PASJ...27..237O,
  1977A&A....58...41A}. Mixed modes and the associated phenomena  of
avoided crossing and mode bumping ---where the periods approach quite
closely without actually crossing--- have been extensively  studied in
the context of sub-giants and red giant pulsating stars  \citep[see,
  e.g.][]{2010Ap&SS.328...51C,2010Ap&SS.328..259D}. In the figure, the  
frequency range of
mixed modes is enclosed with two dashed  lines. Nodes in the radial
eigenfunction for such modes may occur in both the $p$- and $g$-mode
propagation cavities.

The propagation diagram corresponding to the hotter template model
($T_{\rm eff}\sim 9500$ K) is displayed in Fig. \ref{figure_03}.
Although the chemical profiles have not changed as compared  with the
cooler template model, the Brunt-V\"ais\"al\"a frequency  exhibits
changes due to the higher effective temperature.  Specifically,  $N^2$
is larger at the outer regions, pushing  the frequencies of the $p$
modes and some mixed modes to higher values  (shorter periods). In
addition, this model has two thin convective regions in the outer
parts of the star, in contrast to  the single and very wide convection zone
characterizing the cooler template  model (see Fig. \ref{figure_02}).

\subsubsection{Excitation mechanism, driving and damping processes}

The normalized growth rate $\eta$ ($\equiv -\Im(\sigma)/ \Re(\sigma)$,  
where $\Re(\sigma)$ and $\Im(\sigma)$ are
the real and the  imaginary part, respectively, of the complex
eigenfrequency  $\sigma$) in terms of pulsation periods $\Pi$ for
overstable $\ell= 1$ modes corresponding to the cool template 
model ($T_{\rm eff}\sim 7400$ K) is shown in the upper 
panel of Fig. \ref{figure_04}. The lower panel shows the 
results for the hot template model ($T_{\rm eff}\sim 9500$ K).
A value of $\eta > 0$ ($\eta < 0$) implies unstable (stable) modes. 
We emphasize with red dots the unstable modes. The nodes of unstable 
modes are highlighted in Figs. \ref{figure_02} and \ref{figure_03} 
with a ``x'' symbol (in magenta) and connected with thin lines.
In Fig. \ref{figure_04}, the vertical dashed lines 
enclose the period interval of mixed $p-g$ modes. 
As can be seen from the figure, the set of unstable modes for both 
template  models include some 
``pure'' $g$ modes and some $g$ mixed modes\footnote{We 
classify as ``$g$ mixed modes'' those mixed modes 
that have most of their nodes laying in the $g$-mode propagation cavity, 
and as ``$p$ mixed modes'' those mixed modes having most of their 
nodes in the $p$-mode propagation cavity.}. No pure $p$ mode nor $p$ mixed mode 
is excited in the cool template model, although a $p$ mixed
mode (that having $k= 1$) is driven in the hot template model. 
According to Fig. \ref{figure_04}, 
within a given  band of unstable modes, the excitation  is  markedly  
stronger (large values of $\eta$) for the lowest-order $g$ mixed modes, 
characterized  by  short periods (high frequencies).  
We note that the band of excited modes in the cool template model 
($800\ {\rm s} \lesssim \Pi \lesssim 3150$ s) is substantially wider 
than that of the hot template model 
($400\ {\rm s} \lesssim \Pi \lesssim 1400$ s).

\begin{figure} 
\begin{center}
\includegraphics[clip,width=8.5 cm]{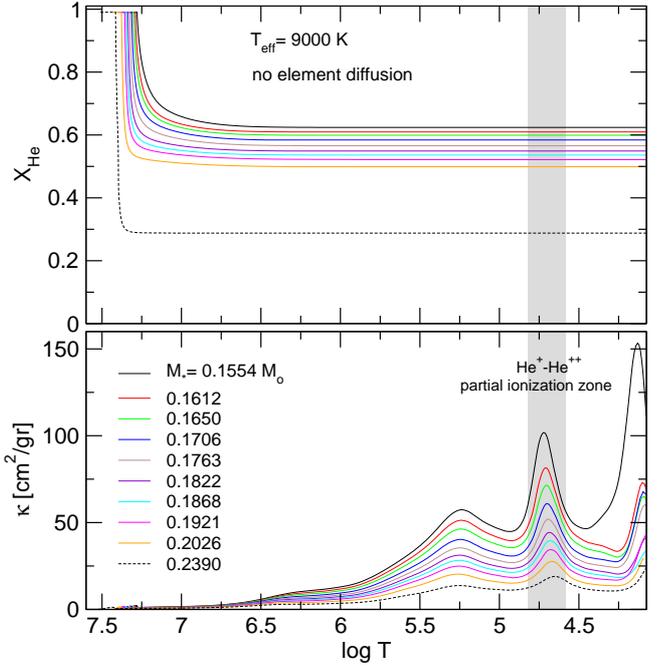} 
\caption{The He abundance (upper panel) and the Rosseland opacity 
(lower panel) in terms of 
the logarithm of the temperature, corresponding to models
at $T_{\rm eff} \sim 9000$ K and different stellar masses indicated
in the plot. The vertical gray strip is 
the location of the region of mode driving ($\log T \sim 4.7$).}
\label{figure_06} 
\end{center}
\end{figure}

In order to investigate the details of the damping/driving processes 
and to identify the excitation mechanism responsible for
the destabilization of modes in our pre-WD models, 
we selected two representative pulsation modes of
each template model. The results described below are also typical 
of all other modes of interest. Specifically, we choose the 
unstable $g$ mixed mode with $k= 5$ and the stable $p$ mixed mode 
with $k= 5$ for both template models. These modes are marked with arrows 
in Fig. \ref{figure_04}. In the left panels of Fig. \ref{figure_05} we 
depict the differential work function $dW/dr$ and the running work integral
$W$ \citep[computed as in][]{1993ApJ...418..855L} 
for the unstable $k= 5$ $g$ mixed mode (stable $k= 5$ $p$ mixed mode), 
characterized by $\Pi= 1091$ s, $\eta= 6.7 \times 10^{-6}$ ($\Pi= 604.7$ s, 
$\eta= -1.03 \times 10^{-5}$) 
in the upper (lower) panel, corresponding to our cool template model
($T_{\rm eff} \sim 7400$ K). The results for the hot template 
model ($T_{\rm eff} \sim 9500$ K) are displayed in the right panels,
being in this case $\Pi= 810.4$ s and $\eta= 3.9 \times 10^{-7}$ for 
the unstable $k= 5$ $g$ mixed mode, and  $\Pi= 330.9$ s and 
$\eta= -1.5 \times 10^{-3}$ for the stable $k= 5$ $p$ mixed mode. 
The scales for $dW/dr$ and $W$ are arbitrary.
Also shown is the  Rosseland opacity ($\kappa$) and its derivatives 
[$\kappa_{\rm  T}+\kappa_{\rho}/(\Gamma_3-1)$],  and the  
logarithm of  the thermal timescale ($\tau_{\rm  th}$).  
We restrict the figures  to the envelope
region of  the models, where the  main driving and  damping processes
occurs. 

Let us first concentrate on the  hot template model (right panels 
of Fig. \ref{figure_05}). The region  
that destabilizes  the modes  (where $dW/dr  > 0$)  is clearly
associated with the bump in the  opacity due to the He$^+$ partial 
ionization zone ($\log T \sim 4.7$), that involves the second ionization of He 
(He$^+ \leftrightarrow$ He$^{++} + e^-$). Note the presence of a 
convection zone at that region (gray area in the Figure),  centered at
$-\log q  \sim 9.7$ $[q \equiv (1-M_r/M_{\star})]$. 
The thermal timescale reaches values in the range 
$\sim 150-1500$ s at the driving  region, in agreement with the 
period interval of unstable modes for this template model ($400-1400$ s). 
In the driving region the quantity $\kappa_{\rm  T}+\kappa_{\rho}/(\Gamma_3-1)$  
is increasing outwards, in agreement
with the well  known necessary condition for mode excitation
\citep{1989nos..book.....U}.  For the unstable $g$ mixed mode, the  
contributions to driving at $-\log q$ from $\sim  9.1$ to $\sim 10.6$ 
(where $dW/dr > 0$) largely overcome the damping  effects at  
$-\log q \lesssim  9$ 
(where $dW/dr < 0$), as reflected by the fact that  $W > 0$ at the 
surface, and therefore the mode is globally excited. In contrast, the strong 
damping experienced by the $p$ mode  (denoted by negative values 
of  $dW/dr$ at $8 \lesssim -\log q \lesssim 9.3$ and 
$-\log q \gtrsim 9.8$) makes this mode globally stable. Note that there 
exists a second, external convective zone, that corresponds 
to the partial ionization region where both the ionization of 
neutral H (H$ \leftrightarrow$ H$^+ + e^-$) and the first 
ionization of He (He $\leftrightarrow$ He$^+ + e^-$) occurs 
($\log T \sim 4.15-4.42$). Since it is located very close to 
the stellar surface, 
the presence of these partial ionization zones is 
completely irrelevant in destabilizing modes. 

\begin{figure} 
\begin{center}
\includegraphics[clip,width=9 cm]{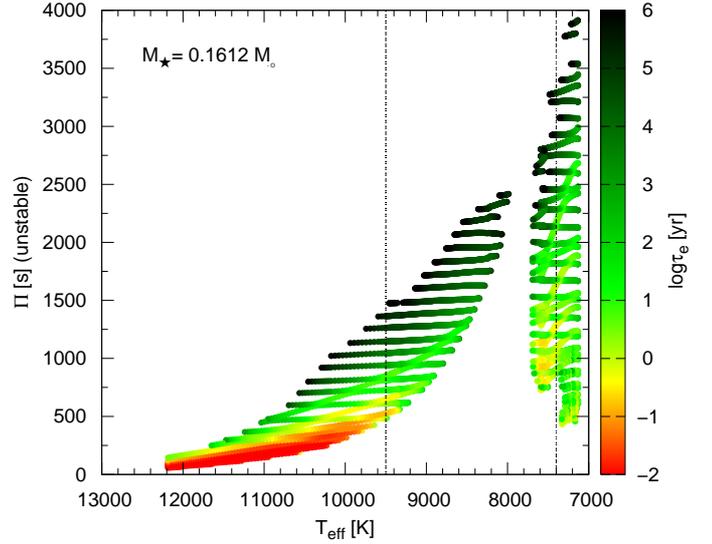} 
\caption{Unstable $\ell= 1$ mode periods ($\Pi$) 
in terms of the effective temperature, corresponding to the pre-WD model 
sequence with $M_{\star}= 0.1612 M_{\sun}$. Color coding indicates the 
value of the logarithm of the $e$-folding time ($\tau_{\rm e}$) of 
each unstable mode (right scale). The vertical lines
indicate the effective temperatures of the template models analyzed 
in Figs. \ref{figure_02}, \ref{figure_03}, \ref{figure_04}, and 
\ref{figure_05}.}
\label{figure_07} 
\end{center}
\end{figure}

We turn now to the case of the cool template model (left panels 
of Fig. \ref{figure_05}). At variance with the hot template model, 
in this case there is a single and wide outer convection zone, 
that extends from $-\log q \sim 6.6$ outwards. 
The opacity is characterized by three bumps,
one is located at  $-\log q \sim 7.9$ and associated with 
the second ionization of He, another at $-\log q \sim 9$ 
corresponding to the first ionization of He, 
and the third is at $-\log q \sim 9.5$ and associated with 
the ionization of neutral H. Note that, since the regions of 
the ionization of H and the first ionization of He are located 
deeper in the star as compared with 
the case of the hot template model, they now play a role in the 
destabilization of modes. Indeed, the three bumps in $\kappa$ are 
responsible for the strong driving of the unstable $g$ mixed mode,
as indicated by the positive values of the differential work function
at those regions, that largely exceeds 
the damping at $-\log q \sim 8.4$ and to a less extent at  
$-\log q \sim 9.4$ and $\sim 9.9$. For the stable $p$ mixed mode,
the driving and damping regions are located at almost 
the same places in the star as for the unstable $g$ mixed mode,
i.e., the function $dW/dr$ exhibits roughly the same shape. 
What makes the difference, 
however, is the very strong damping experienced by this mode
at the surface regions ($-\log q \gtrsim 9.7$), where $dW/dr$
adopts large and negative values, which render the mode globally stable.

In summary, we identify the $\kappa-\gamma$ mechanism  associated with
the opacity bump due to partial ionization   of He$^+-$ He$^{++}$
centered at $T \sim 5 \times 10^4$ K as a  destabilizing agent for
the unstable pulsation modes of our hot pre-WD models, which  are
represented by the template model at $T_{\rm eff} \sim 9\,500$ K
(right upper panel of Fig. \ref{figure_05}).  This is in agreement
with the predictions of the radial-mode instability  results of
\citet{2013MNRAS.435..885J} (see \S \ref{comparison}). Note that the
extension (in $T_{\rm eff}$) of the instability domain is  sensitive
to the abundance of He at the driving region, which in turn, depends
on the stellar mass and possibly to the detailed previous 
binary evolution. In Fig. \ref{figure_06} we show the He abundance
(upper panel) and the Rosseland opacity (lower panel) in terms of  the
logarithm of the temperature, corresponding to models at $T_{\rm eff}
\sim 9000$ K and different stellar masses. We  emphasize with a
vertical gray strip the driving region of the  models ($\log T \sim
4.7$). The abundance of He at the driving region  monotonically
decreases with increasing stellar mass (upper panel).  Also,
the density at the driving region diminishes for higher masses.
Thus, because He is less abundant and density is lower for stellar 
models characterized by higher masses, the bump in the  opacity at the
region of the He$^+-$ He$^{++}$ partial ionization zone decreases in
magnitude for increasing stellar mass. This translates in a weaker
capability to destabilize modes. Ultimately, for models  with masses
$M_{\star}/M_{\sun} \geq 0.2390$ and $T_{\rm eff} \sim 9000$ K  
the He abundance  is so low (and the
bump in the opacity has  decreased to such an extent) that there is no
pulsational  instability, and models in that range of masses  do not
exhibit excited modes (see Fig. \ref{figure_01}).

On the other hand, we have found that for the cool pre-WD models,
represented by our template model at $T_{\rm eff} \sim 7400$ K (left
upper panel of Fig. \ref{figure_05}), the $\kappa-\gamma$ mechanism
is efficient to destabilize modes. This time, however, the
$\kappa-\gamma$ mechanism not only works  due to the partial
ionization    of He$^+-$ He$^{++}$ ($T \sim 5 \times 10^4$ K), but
instead also due to the  presence of the bumps in $\kappa$ associated with 
the partial ionization    of He$-$ He$^{+}$ and H$-$ H$^{+}$, located at
$T \sim  2.6\times 10^4$ and  $T \sim  1.4\times 10^4$, respectively. 

\subsubsection{Instability domains}

A more comprehensive view of the pulsational  stability properties of
our models can be accomplished by examining the range of unstable
periods of our template model  for the full range of effective
temperatures. In Fig. \ref{figure_07} we show  the instability domains
of $\ell= 1$ periods in terms of the effective temperature for the
pre-WD model sequence with  $M_{\star}=  0.1612  M_{\sun}$. The
palette of  colors (right scale) indicates the value  of the logarithm
of the $e$-folding time (in years) of each unstable  mode. The
vertical lines correspond to the effective temperatures of the two
template models described in Figs. \ref{figure_02}, \ref{figure_03},
\ref{figure_04},   and \ref{figure_05}.   The $e$-folding time,
defined  as  $\tau_{\rm  e}=  1/|\Im(\sigma)|$,   is an estimate of
the time it  would take a given mode  to reach   amplitudes large
enough as  to be observable. It has  to be   compared with  the
evolutionary timescale  which represents   the time that the
model  spends evolving in the regime  of interest.  In the case
displayed in  Fig. \ref{figure_07}, 
$\tau_{\rm instability} \sim 1.3 \times 10^{9}$ yr,
$\tau_{\rm instability}$ being the time it takes  the $0.1612M_{\sun}$ model to
evolve  from $T_{\rm eff} \sim 7200$ K to $T_{\rm eff} \sim 12\,400$
K (Table \ref{table1}). Note  that 
$\tau_{\rm instability} >> \tau_{\rm e}^{\rm max}\sim 10^6$ yr. Thus,  our
calculations show that there is plenty of time for  the
destabilization of any of the pulsation modes shown in the  Figure,
and to achieve  observable amplitudes while  the star is in the 
instability domain.

Two different and separated domains of instability can be clearly
distinguished in the  Fig. \ref{figure_07}. One of them, at high
temperatures, corresponds  to modes destabilized by the
$\kappa-\gamma$ mechanism associated with the second ionization of
He. The other domain, at lower effective  temperatures, is associated
with the $\kappa-\gamma$ mechanism  in which, in addition to the second
ionization of He,   there are appreciable contributions from the first
ionization of He  and the ionization of H, as we have discussed
before. Notably, there are evident signatures of avoided crossings
in both instability domains.

For the high-$T_{\rm eff}$ domain, among the unstable modes, the
strongest   excitation (that is,   the smallest  $e$-folding time,
emphasized with light red and yellow  zones in the plot) corresponds to
short periods in the range $150 {\rm s} - 600$ s associated with
low-order $p-g$   mixed modes. Intermediate- and high-order $g$ modes
with periods  $\Pi \gtrsim 600$ s are also driven, but with much
longer  $e$-folding times (light and dark green).  The trend of
shorter excited periods for higher effective  temperatures that
characterizes the instability domain in  the $T_{\rm eff}- \Pi$ plane
(see Fig. \ref{figure_07}) can be explained in terms of the precise
location of the driving region in the interior of the model and the
thermal  timescale at that region. Specifically, as the star
contracts and evolves towards higher effective temperature,
the partial ionization zone of He$^+-$ He$^{++}$ shifts toward layers
located  closer to the stellar surface, where the thermal timescale is
shorter.  Thus, as the $T_{\rm eff}$ of the star increases, the
$\kappa-\gamma$ mechanism is able to excite modes with gradually
shorter periods.  Eventually, when the partial ionization zone of
He$^+-$ He$^{++}$ is very close to the stellar surface, it loses the
ability to  destabilize modes, giving place to the hot edge of the
instability  domain.

\begin{figure*} 
\begin{center}
\includegraphics[clip,width=17 cm]{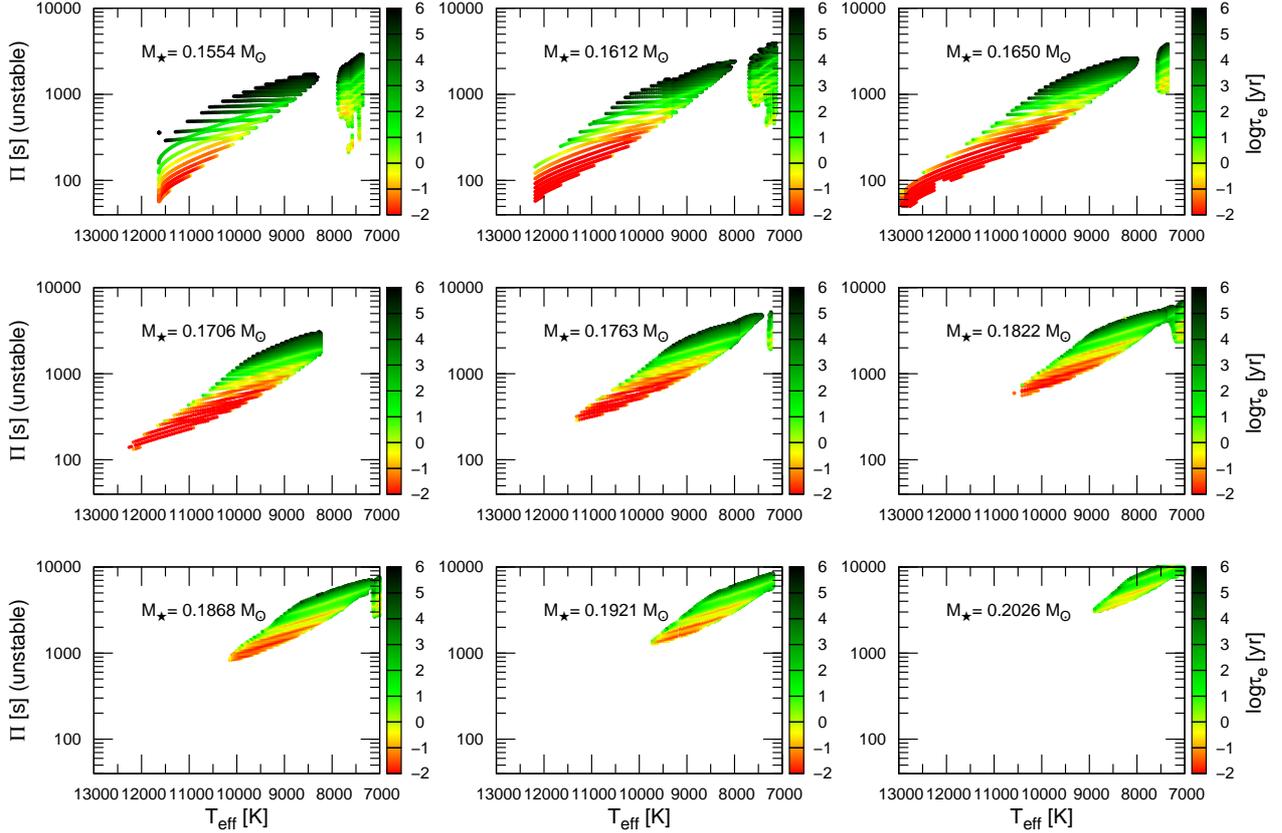} 
\caption{Unstable mode periods ($\Pi$) for $\ell= 1$ in terms of
  the effective  temperature, corresponding  to the  pre-WD model
  sequences with stellar masses $M_{\star}/M_{\sun}= 0.1554, 0.1612, 0.1650, 
  0.1706, 0.1763, 0.1822, 0.1868, 0.1921$, and $0.2026$. 
  Color coding indicates the value of the logarithm  of the
  $e$-folding  time ($\tau_{\rm e}$) of  each  unstable mode  (right
  scale).}
\label{figure_08} 
\end{center}
\end{figure*}

\begin{figure*} 
\begin{center}
\includegraphics[clip,width=17 cm]{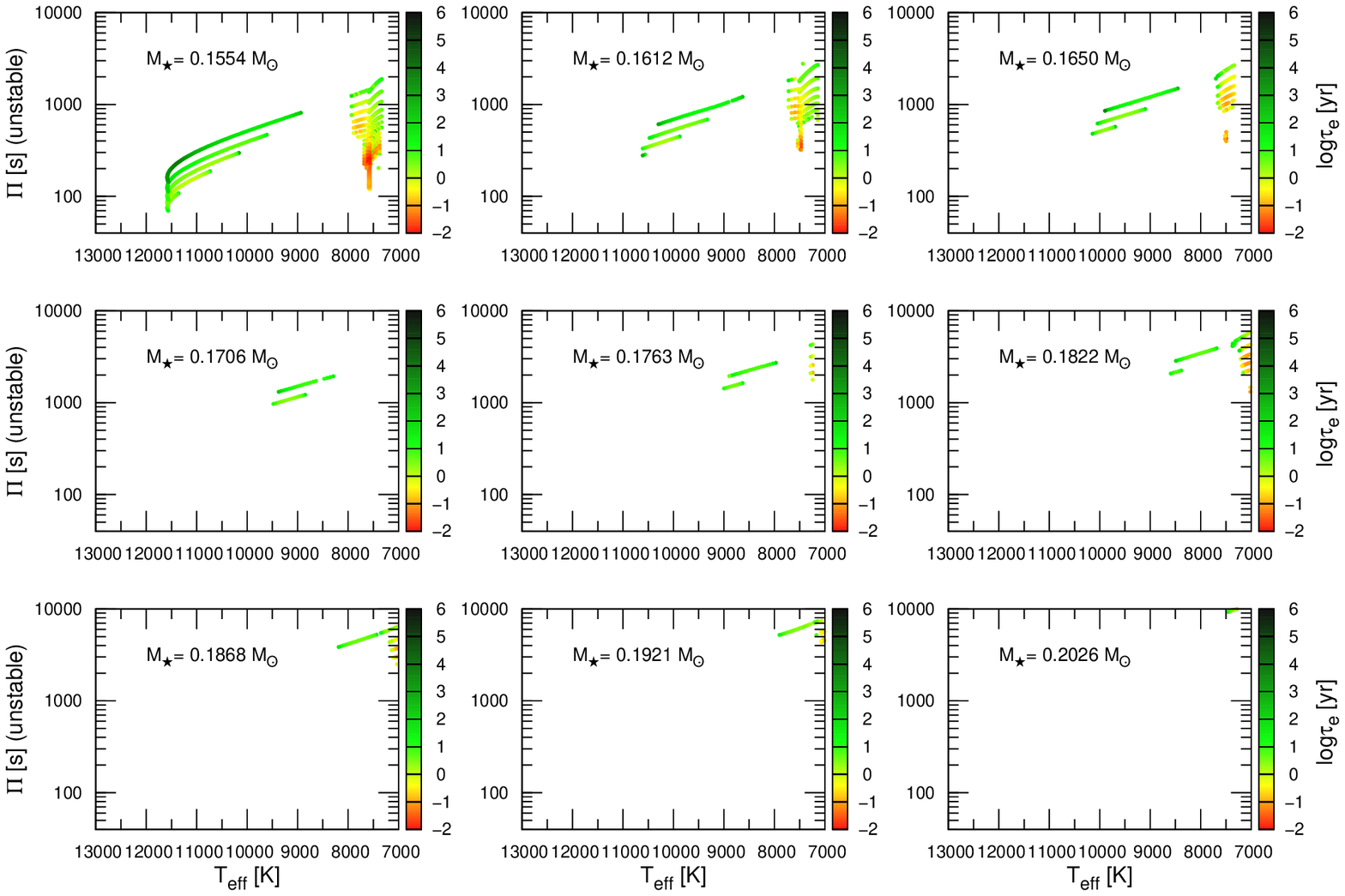} 
\caption{Same as Fig. \ref{figure_08}, but for radial ($\ell= 0$) modes.}
\label{figure_09} 
\end{center}
\end{figure*}

We focus now on the instability domain at lower $T_{\rm eff}$. 
This instability region is  characterized by much wider ranges of excited
modes, as compared to  the  main instability region at higher
$T_{\rm eff}$. Note that there  exists a gap in effective temperature,
separating both instability domains, in which no unstable modes exist. 
As described above, at low
effective temperatures  ($T_{\rm eff} \lesssim 7800$ K), the partial
ionization regions  of H$-$H$^+$ and He$-$He$^{+}$ are far enough from
the stellar  surface such that they contribute substantial driving, 
adding to that from the He$^+-$He$^{++}$ partial ionization
zone.  As a result of this, the driving region is fairly wide, which
translates  to a very broad range of unstable periods. As the star
evolves and heats up,  the opacity bump corresponding to the
He$^+-$He$^{++}$ partial ionization  zone drops significantly, while
the partial ionization region of  H$-$H$^+$ and He$-$He$^{+}$ shifts
towards the stellar surface.  Both effects result in a strong
weakening of the total driving, leading to an instability domain
that is quite restricted in the $T_{\rm eff}-\Pi$
plane or that can even disappear altogether. 
As the  star heats up more, the partial ionization region of
He$^+-$He$^{++}$ eventually becomes efficient enough again as to
destabilize modes, giving origin  to the main instability region, at
high $T_{\rm eff}$, as described before. 

The dependence described above between the periods of unstable $\ell=
1$  modes and the $T_{\rm eff}$ of the models holds for all of our
pre-WD sequences, as is apparent from Fig. \ref{figure_08}.  In
this multiple-panel array we show the unstable modes on the $T_{\rm
  eff}- \Pi$ plane for the evolutionary sequences with masses
$M_{\star}/M_{\sun}= 0.1554, 0.1612, 0.1650, 0.1706, 0.1763,  0.1822,
0.1868, 0.1921$, and $0.2026$. The sequence  with $M_{\star}= 0.2390
M_{\sun}$ (not included in the Figure)  exhibits unstable models with
periods longer than $\sim 13\,000$ s.  More massive model sequences do
not show unstable modes for the  range of effective temperatures
explored ($T_{\rm eff} \gtrsim 6000$ K).   It is important to mention
that, at effective temperatures  higher than $\sim 6\,000$ K, there
exist pulsating stars with modes that appear to be excited either stochastically 
or coherently by near surface convection, an example
being the $\delta$ Scuti  star HD 187547 \citep[$T_{\rm eff}\sim 7500$
  K,  $\log g\sim 3.90$;][]{2011Natur.477..570A,2014ApJ...796..118A}.
So, it is not inconceivable that some pulsation modes detected in
low-mass  pre-WD stars can be driven not by the  $\kappa-\gamma$
mechanism, but by stochastic excitation instead.

In Fig.\,\ref{figure_08}
we have employed a logarithmic scale for the periods in order to
enhance the short-period regime of the  $T_{\rm eff}- \Pi$
diagrams. The figure shows  the presence of the two
instability regions described before  for $M_{\star}/M_{\sun} \leq
0.1868$. We note that both regions are separated for
$M_{\star}/M_{\sun} \leq 0.1763$, but they merge into one for
sequences with larger masses.  In general, both $p$ and $g$ modes are
excited  in our pre-WD models, including modes with mixed
character ($p$ and $g$; see Figs. \ref{figure_02}, \ref{figure_03}
and \ref{figure_04}).  The longest excited periods reach values in
excess of $\sim 15\,000$ s for the highest-mass sequence showing
unstable modes  ($M_{\star}/M_{\sun}= 0.2026$), and this limit
drastically decreases to $\sim 3000$ s for the less massive  sequence
($M_{\star}/M_{\sun}= 0.1554$) considered.   The shortest excited
periods, in turn, range from $\sim 50$ s for $M_{\star}/M_{\sun}=
0.1554$,  to $\sim 3000$ s  for $M_{\star}/M_{\sun}= 0.2026$.  So, the
longest and shortest excited periods are  larger for higher
$M_{\star}$. We note that for the case of $\ell= 2$ modes, the longest and 
shortest periods are somewhat shorter than for $\ell= 1$ modes.
We note that, in the case of the sequences  with
$M_{\star}/M_{\sun}= 0.1554, 0.1650$ and $0.1612$,  the stellar models
still display excited modes even after reaching  their maximum
effective temperature (see Fig. \ref{figure_01})  and entering their
cooling tracks. For clarity, we have not included in
Fig. \ref{figure_08} the unstable periods associated with these
stages. Regarding the strength of the mode instability, we found that,
for each mass considered, the modes with shorter periods are the 
most strongly excited, with $e$-folding times as short as
$\tau_{\rm e} \sim 0.01-1$ yr). In contrast, modes
with longer periods are much more weakly destabilized,
and $e$-folding times can be as large as 
$\tau_{\rm e} \sim 10 - 10^6$ yr, although still
much shorter than the  time it takes models to cross
the instability domains, i.e.  $\tau_{\rm e} \ll 
\tau_{\rm instability}$ (see Table \ref{table1}). 

Fig. \ref{figure_09} is the counterpart of Fig. \ref{figure_08} but
for  radial ($\ell= 0$) modes. It may be noted that, although the
trends are the same as for nonradial modes with $\ell= 1$,  in the
case of radial modes there are many fewer modes excited.  This trend
is more pronounced for higher masses. In the case of  the high-$T_{\rm
  eff}$ instability domain, radial modes with $k= 0,1,2,3,4$ are
excited for the sequence with $M_{\star}= 0.1554  M_{\sun}$, whereas
for sequences more massive than $0.1822 M_{\sun}$, only the
fundamental radial ($k= 0$) mode is excited. On the other hand,  in
the case of  the low-$T_{\rm eff}$ instability domain, a modestly
higher number of  radial modes  are excited. Finally, we note that
the blue edge of the radial-mode instability domain is substantially cooler 
than that of nonradial $\ell= 1$ modes (see Fig. \ref{figure_01}).

\subsubsection{Comparison with the \citet{2013MNRAS.435..885J}'s results}
\label{comparison}

\citet{2013MNRAS.435..885J} have performed a detailed 
exploration of radial-mode ($\ell= 0$) pulsation stability  
of low-mass pre-WD stars across a range of composition, effective 
temperature and luminosity. They present their results in terms of the 
instability domains in the $\log T_{\rm eff} - \log[(L_{\star}/L_{\sun})/
(M_{\star}/M_{\sun})]$ 
space. In order to make a comparison with our own 
results, we transform their results to the $T_{\rm eff} - \log g$ plane. 
\citet{2013MNRAS.435..885J} employ a set of 
static envelope models with uniform compositions of 
$X_{\rm H}= 0.90, 0.75, 0.50, 0.25, 0.01$ and $0.002$ and a basic metal 
mass fraction $Z = 0.02$. This implies He fractional abundances
of  $X_{\rm He}= 0.08, 0.23, 0.48, 0.73, 0.97$ and $0.978$.
Since our evolutionary models are characterized by 
$0.49 \lesssim  X_{\rm He} \lesssim 0.62$ in the mass range 
$0.1554 \lesssim M_{\star}/M_{\sun} \lesssim 0.2026$  
(see Fig. \ref{figure_06}), we can compare our results with 
those of \citet{2013MNRAS.435..885J}
for the case $X_{\rm H}= 0.50$ ($X_{\rm He}= 0.48$). 

In Fig. \ref{figure_10}
we show the same $T_{\rm eff} - \log g$ diagram presented in Fig. \ref{figure_01}
but this time including a gray area that represents the results of 
\citet{2013MNRAS.435..885J} for their radial-mode  instability  
domain for static envelope models with a 
H fractional mass of $X_{\rm H}= 0.50$ ($X_{\rm He}= 0.48$), 
as depicted in the upper right panel of their Figure 1.
The region enclosed by the instability boundaries includes 
models with at least one unstable radial mode. Note that the instability domain
predicted by our calculations (both for radial and nonradial modes) 
is contained within the instability domain of \citet{2013MNRAS.435..885J}. 
Note, however, that our radial-mode blue edge  is  cooler 
($\sim 800$ K) than the blue edge of the 
\citet{2013MNRAS.435..885J} instability domain. Notably, it is our  
nonradial-mode blue 
edge ($\ell= 1, 2$) which is 
in excellent agreement with the radial-mode blue edge of those
authors. The discrepancy between 
our $\ell= 0$ blue edge and that of \citet{2013MNRAS.435..885J}
could be explained, in part,  on the basis that our models have a He 
abundance at the driving region which is dependent on the stellar mass, 
while the models of \citet{2013MNRAS.435..885J} 
have a fixed He abundance ($X_{\rm He}= 0.48$).
On the other hand, there is a  notable agreement between the two sets of 
calculations 
in relation to the unstable region where the partial ionization of 
H$-$H$^+$ and He$-$He$^{+}$ contributes to the destabilization of 
modes. Indeed, in Fig. \ref{figure_10} the hot boundary of the instability 
``finger'' at $7800 \lesssim T_{\rm eff} \lesssim 9000$ K and 
$4.5 \lesssim \log g\lesssim 6$  
of \citet{2013MNRAS.435..885J} nearly matches the extension of 
the dot-dashed  lines corresponding to our computations.

In summary, given that both sets of computations were performed 
independently using different evolutionary and pulsation codes,  and 
very distinct stellar models, and in spite of the discrepancy in the 
location of the radial-mode blue edges, the agreement between 
the results of \citet{2013MNRAS.435..885J}  and ours  is 
encouraging.

\subsection{Computations with element diffusion}
\label{withdiff}

\begin{figure} 
\begin{center}
\includegraphics[clip,width=9 cm]{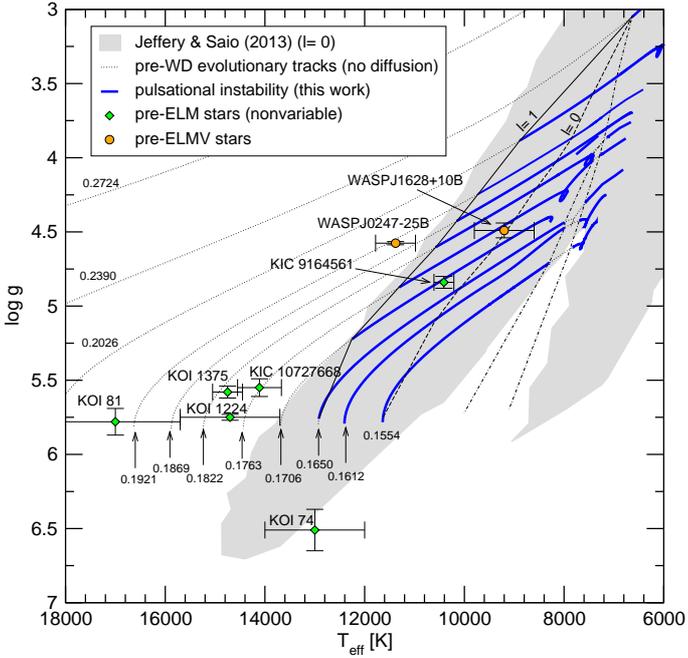} 
\caption{Same as Fig. \ref{figure_01}, but including the radial 
mode ($\ell= 0$) instability  
domain (gray area) computed by \citet{2013MNRAS.435..885J} for 
the the case of 
static envelope models with a H fractional mass 
of $X_{\rm H}= 0.50$ ($X_{\rm He}= 0.48$) (upper right panel 
of their Fig. 1).} 
\label{figure_10} 
\end{center}
\end{figure}

\begin{figure} 
\begin{center}
\includegraphics[clip,width=9 cm]{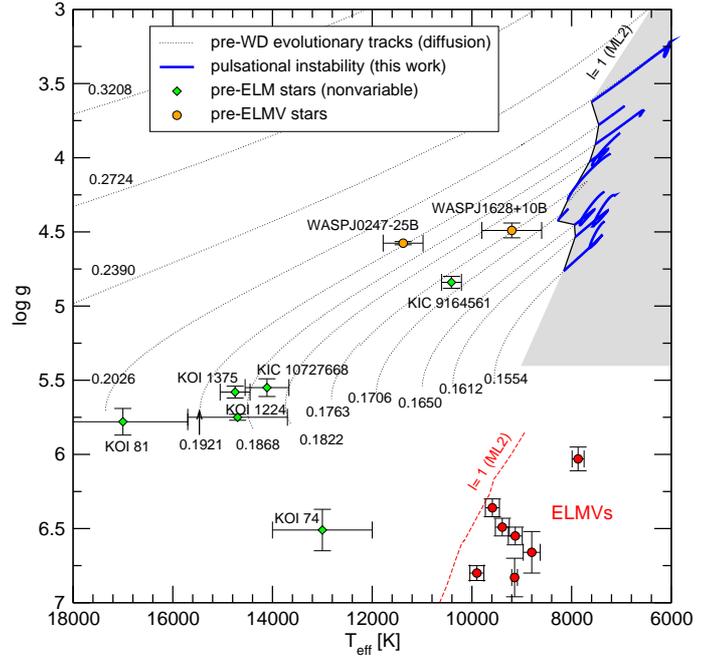} 
\caption{Same as Fig. \ref{figure_01}, but for the case in which 
the low-mass He-core pre-WD models are computed by taking into 
account  element diffusion. The evolutionary tracks are marked with 
dotted curves.} 
\label{figure_11} 
\end{center}
\end{figure}

We have redone all of our evolutionary and stability pulsation computations 
by taking into account the action of element diffusion, as in
\citet{2013A&A...557A..19A}. In Fig. \ref{figure_11} we present a
$T_{\rm eff} - \log g$ plane showing the resulting low-mass He-core 
pre-WD evolutionary tracks (dotted curves) computed considering 
element diffusion. The evolutionary tracks 
with element diffusion are substantially different to the case in which 
this process is neglected. In particular, diffusion tracks  
are displaced to lower $T_{\rm eff}$ and $\log g$ as compared with 
non-diffusion tracks, as it can be realized by comparing
Fig. \ref{figure_11} with Fig. \ref{figure_01} (see also Fig. \ref{figure_10}). 
This is because, for a given surface luminosity value, models with
element diffusion have lower effective temperature and thus larger radii 
owing to the fact that H diffuses upwards during the course of evolution
\citep[see][]{2000MNRAS.317..952A,2001ApJ...554.1110A}.

Fig. \ref{figure_11} demonstrates emphatically that when element
diffusion is taken into account in low-mass He-core pre-WD models,
the pulsational instability region is much more restricted than in the
case of non diffusion models.  This can be understood as follows.
When the models, which are evolving towards higher effective
temperatures,   are still cool enough, there is strong driving due to
the $\kappa-\gamma$ mechanism acting at the three partial  ionization
regions present in the stellar envelope: He$^+-$He$^{++}$ ($\log T
\sim 4.7$), He$-$He$^{+}$ ($\log T \sim 4.42$), and  H$-$H$^+$ ($\log
T \sim 4.15$). However, as element diffusion  operates, H begins to
migrate to the surface and  He starts to sink
toward deeper layers. The increasing deficit of He  strongly weakens
the driving due to the He$^+-$He$^{++}$ and He$-$He$^{+}$  partial
ionization zones, while the driving due to the  H$-$H$^+$ partial
ionization zone becomes progressively important.  When $X_{\rm He}$
drops  below a certain critical value at the  driving regions,
pulsational instabilities are due exclusively to  the H$-$H$^+$
partial ionization zone. While this zone is located  far enough from
the surface, the driving will continue due to the $\kappa-\gamma$
mechanism. However, this partial ionization region migrates towards
the stellar surface as the star  ---while evolving to higher $T_{\rm
  eff}$-- heats up.  Consequently, the efficiency of driving is
weakened, until the instability  ceases. The blue edge appears at
that stage. All this process  involves a rather discrete range in
$T_{\rm eff}$ ($\sim 1000$ K),  and this is the reason  for which the
instability domain is so narrow.  This is clearly shown in
Fig. \ref{figure_11} by means of thick  blue lines that mark the
portions of the evolutionary tracks corresponding to stellar models
having unstable modes. For illustrative  purposes, we extrapolate the
instability region (gray area) to lower and higher  gravities, and
to lower effective temperatures. 

\begin{figure} 
\begin{center}
\includegraphics[clip,width=8.5 cm]{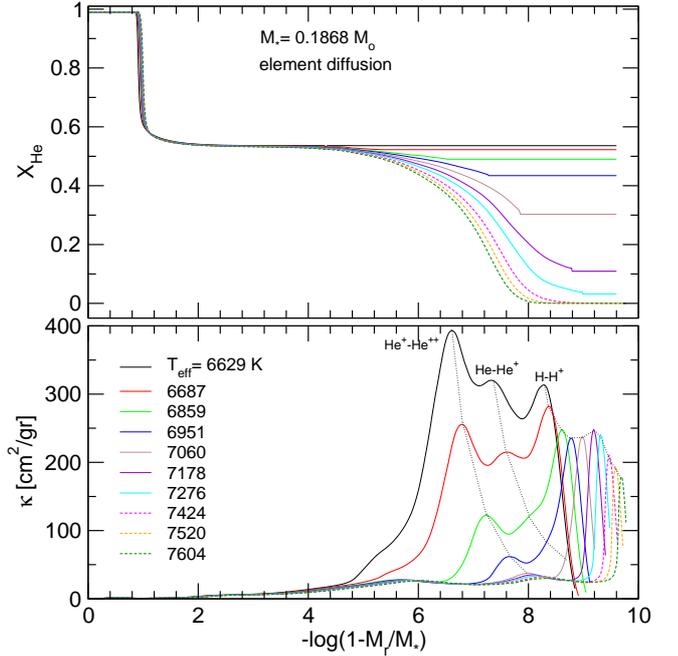} 
\caption{The He abundance (upper panel) and the Rosseland opacity 
(lower panel) in terms of the outer mass fraction coordinate, 
corresponding to models with $M_{\star}= 0.1868 M_{\sun}$ with 
element diffusion and different (increasing) values of $T_{\rm eff}$, 
which are marked in Fig. \ref{figure_13}. Solid curves correspond
to models in which there is pulsation instability, while
dashed curves are associated with models in which instability has 
ceased. The tops of the bumps
in the opacity due to the He$^+-$He$^{++}$, 
He$-$He$^{+}$, and H$-$H$^+$ partial ionization regions are
connected with thin dotted lines, to show their evolution.}
\label{figure_12} 
\end{center}
\end{figure}

\begin{figure} 
\begin{center}
\includegraphics[clip,width=9 cm]{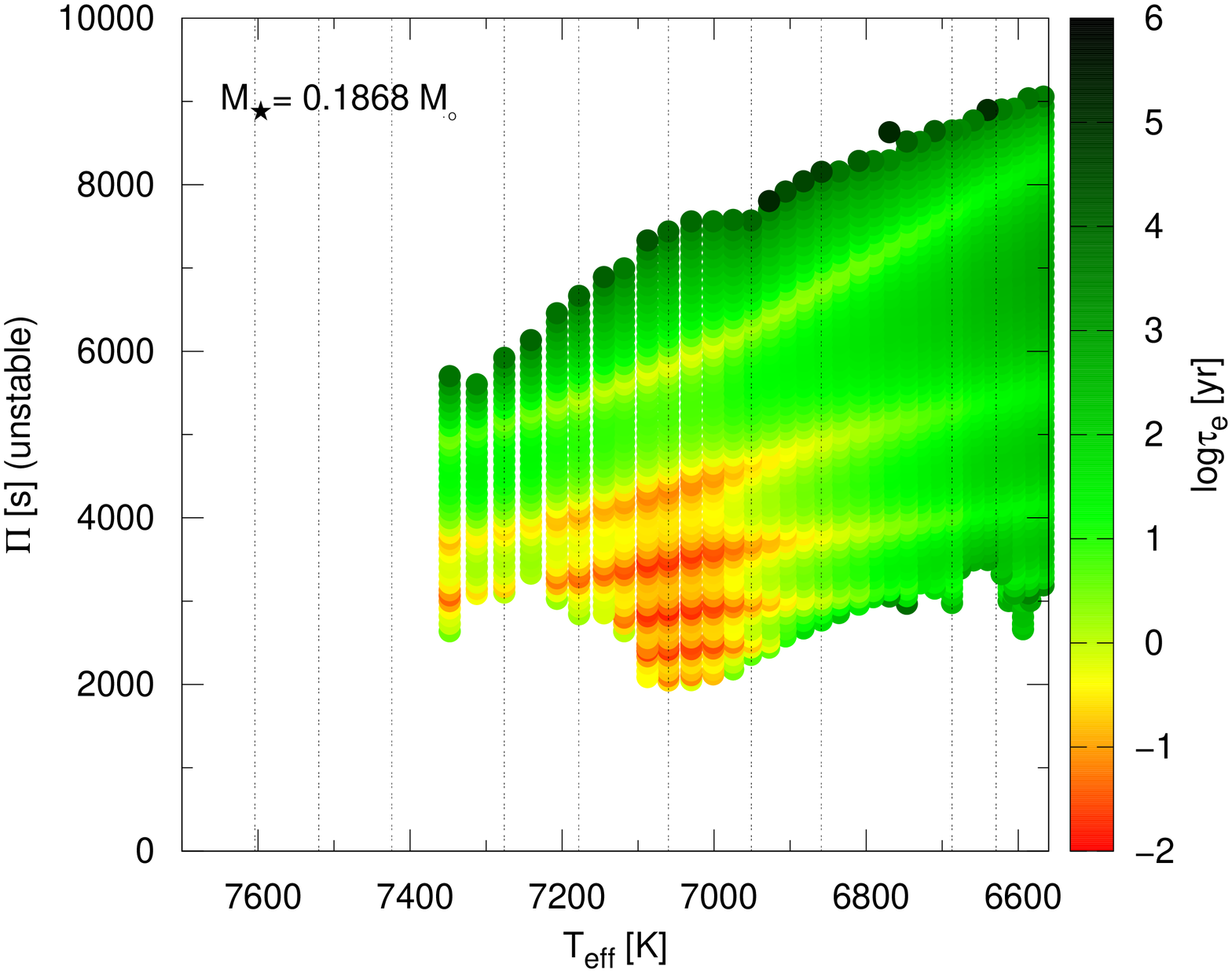} 
\caption{Unstable $\ell= 1$ mode periods ($\Pi$) 
in terms of the effective temperature, corresponding to the pre-WD model 
sequence with $M_{\star}= 0.1868 M_{\sun}$. Color coding indicates the 
value of the logarithm of the $e$-folding time ($\tau_{\rm e}$) of 
each unstable mode (right scale). The vertical dashed lines
indicate the effective temperatures of the models analyzed 
in Fig. \ref{figure_12}.}
\label{figure_13} 
\end{center}
\end{figure}

In the upper panel  of Fig. \ref{figure_12} 
we show the He abundance in terms of the 
outer mass fraction for different effective temperatures 
during the evolution of $0.1868 M_{\sun}$ pre-WD 
models when element diffusion is taken into account. The lower 
panel shows the Rosseland opacity for 
the same evolutionary stages. As might be expected, 
the contribution of the 
He$^+-$He$^{++}$ and He$-$He$^{+}$ $\kappa$ bumps strongly 
diminishes as the abundance of He drops at the locations 
of the corresponding partial ionization zones, 
to the point that for $T_{\rm eff} \sim 7060$ K these bumps have 
vanished. At the same time, the $\kappa$ bump due to the 
H$-$H$^{+}$ partial ionization zone maintains its value almost 
constant, and remains  the only source of mode 
excitation. For higher effective temperatures, this bump moves
towards the stellar surface, gradually losing its ability to 
excite modes through the $\kappa-\gamma$ mechanism. 
For $T_{\rm eff}\gtrsim 7400$ K, pulsational excitation by this partial 
ionization region ceases (dashed curves) and all the modes become 
pulsationally stable. All this process also can be appreciated 
in the $\Pi-T_{\rm eff}$ plane, as depicted in Fig. \ref{figure_13},
in which we show the evolution of the $\ell= 1$ unstable mode periods
corresponding to the model sequence with $M_{\star}= 
0.1868 M_{\sun}$. Vertical dashed lines indicate the 
$T_{\rm eff}$ values considered in Fig. \ref{figure_12}.
For $T_{\rm eff} \gtrsim 7000$ K, the unstable modes 
shown in the figure are destabilized by the  H$-$H$^{+}$ partial 
ionization zone solely. Interestingly enough, modes
with periods in the range $2000 \lesssim \Pi \lesssim 5000$ s are strongly
destabilized, as can be appreciated from the light red and yellow zones
associated with very short $e$-folding times. 

\section{Comparison with the observed pre-ELMVs}
\label{observed}

At present, only two variable stars identified as pulsating pre-ELM WD
objects have been detected.  They are plotted  in
Figs. \ref{figure_01}, \ref{figure_10}, and \ref{figure_11}.   We
  include in these figures six nonvariable pre-ELM WDs observed in the
  \emph{Kepler} mission field.  The pulsating stars are WASP
J0247$-$25B,  with spectroscopic parameters of $T_{\rm  eff}=
11\,380 \pm 400$  K and  $\log  g= 4.576\pm 0.011$
\citep{2013Natur.498..463M}, and WASP J1628$+$10B, characterized by
$T_{\rm eff}= 9\,200\pm600$ K,  $\log g= 4.49\pm0.05$
\citep{2014MNRAS.444..208M}\footnote{We note that the primary stars in
  both cases are  A-type stars (showing $\delta$ Scuti
  pulsations), while in our  computations we consider that the primary
  star of the initial binary system  is a neutron star. We emphasize
  that we treat the neutron star  essentially as a tool for stripping
  the envelope from the pre-WD star \citep{2013A&A...557A..19A}.}.

From Sect. \ref{withdiff}, we note that our  theoretical computations
with  element diffusion do not account for any of the observed
pulsating  pre-ELM WD stars. We conclude that, on the basis of our
nonadiabatic pulsation models,  element diffusion does not operate in
low-mass  pre-WD stars, at least in the frame of the binary
  evolution model  assumed in our study for the evolutionary history
  of progenitor stars.  Several factors could play a role in
preventing (or undermining)  the effects of element diffusion,
including stellar winds  \citep{2000A&A...359.1042U} and/or stellar
rotation \citep{1971MNRAS.152...47S}.  As interesting as these topics
are in this  context, their exploration is  beyond the scope of this
investigation and  they will be addressed in future studies.

The remainder of the discussion in this Section is focused on the
comparison  between the observed properties of the target stars and
the predictions emerging from our stability analysis based on 
non-diffusion models that have been explored at length 
in Sect. \ref{withoutdiff}.
Fig. \ref{figure_01} shows that for these models, 
WASP J1628$+$10B is located within
the theoretical instability domains.  
As for WASP J0247$-$25B, it is hotter than the blue 
edges ($\ell= 0, 1, 2$) of the instability domain 
in the  $T_{\rm eff}-\log g$ plane (Fig. \ref{figure_01}). 
On the other hand, of the six nonvariable pre-ELM WDs considered, 
five of them fall outside the theoretical instability strip, and only one
(KIC 9164561) is oddly located within the nonradial instability region.

Below, we first explore the excited period spectrum of
WASP J1628$+$10B in the context of our non-diffusion stability
computations. Then we consider the possibility of whether a
higher He abundance at the driving regions in our models would
extend the blue edge of the instability domain in order to include
WASP J0247$-$25B. Finally, we estimate how small the abundance 
of He at the driving region of KIC 9164561 must be in order to inhibit 
the pulsations in this star. 

\subsection{WASP J1628$+$10B}

\begin{figure} 
\begin{center}
\includegraphics[clip,width=8.5 cm]{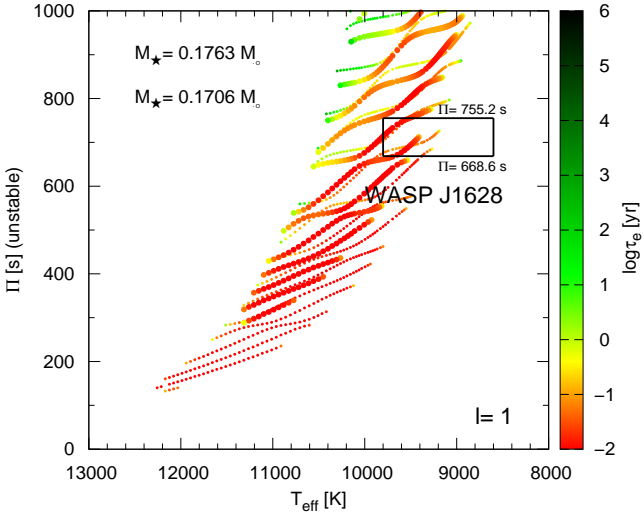} 
\caption{The periods of unstable $\ell= 1$ modes in terms of the effective
temperature, with the palette of colors (right scale) indicating the value
of the logarithm of the $e$-folding time (in years), 
corresponding to the sequences with $M_{\star}= 0.1706 M_{\sun}$
(small dots) and $M_{\star}= 0.1763 M_{\sun}$ (large dots) 
neglecting element diffusion. Also shown are the pulsation periods of
the pre-ELM WD star WASP J1628$+$10B (horizontal segments).}
\label{figure_14} 
\end{center}
\end{figure}

According to \citet{2014MNRAS.444..208M}, this star exhibits
pulsations with periods at $668.6$ s and  $755.2$ s. The effective
temperature and gravity situate WASP J1628$+$10B within the theoretical
instability domain derived in this work
(Fig. \ref{figure_01}). According to its location  in the $T_{\rm
  eff}-\log g$ diagram, we can compare the periods observed in this
star with the ranges of excited periods for the sequences  with
$M_{\star}= 0.1706 M_{\sun}$ and $M_{\star}= 0.1763 M_{\sun}$.  In
Fig. \ref{figure_14} we show the periods of unstable $\ell= 1$  modes
for these two sequences in terms of the effective temperature, along
with the pulsation periods of the pre-ELM WD star WASP J1628$+$10B.
Notably, our theoretical computations are in good agreement  with the
observations. In particular, we obtain unstable modes with periods
that closely match the periods measured in  WASP J1628$+$10B at the
right  effective temperature range. We note  that, according to 
our theoretical
computations, the modes associated with the periods of WASP J1628$+$10B
are among the most unstable modes, with very short $e$-folding times
($\tau_{\rm e} \sim 0.01-1$ yr).

We have not  considered radial modes in our comparison, since our
computations for these stellar masses predict only the
fundamental  ($k= 0$) and first overtone ($k= 1$) radial modes to be
unstable for these stellar masses (see Fig. \ref{figure_09}),  
with periods long in excess ($\sim 1000-3000$ s) and  therefore
inconsistent with the periods observed in this star. 

\subsection{WASP J0247$-$25B}

As mentioned, the pulsating pre-ELM WD star WASP J0247$-$25B
(with periods at $380.95$ s, $405.82$ s, and $420.64$ s) 
falls outside the pulsation instability domain computed in this 
work for non-diffusion models.
This may be indicating that the He abundance in the driving 
region of the 
star is larger than our evolutionary computations 
predict.  Also, part of this discrepancy could be attributed 
to the FC treatment employed in this work, 
that might be introducing uncertainties in the precise 
location of our blue edges.   
Here, we explore how large the He abundance should  be 
to get pulsational instability at the right effective temperature 
and in the range of periods displayed by this star. 
With this aim, we evolve our pre-WD models up to the 
$T_{\rm eff}$ of the star, and then we artificially modify 
$X_{\rm He}$ at the location of the opacity bump due to 
the He$^+-$He$^{++}$ partial ionization zone. We perform
nonadiabatic computations for increasing values of 
$X_{\rm He}$, starting with the canonical value predicted 
by the previous evolution. We restrict our analysis to nonradial 
$\ell= 1$ pulsation modes.

\begin{figure} 
\begin{center}
\includegraphics[clip,width=8.5 cm]{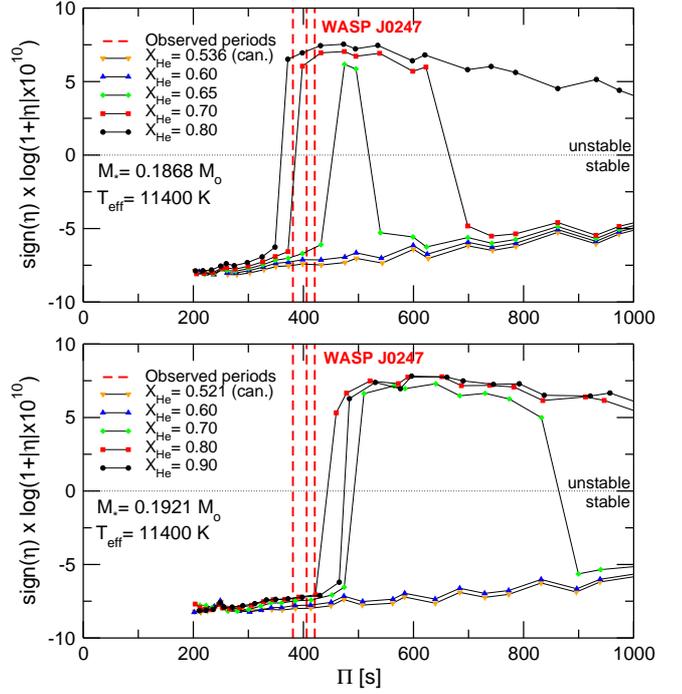} 
\caption{Normalized growth rates $\eta$ for $\ell= 1$  modes 
in terms of the pulsation periods for  $0.1868 M_{\sun}$ pre-WD  
models  at $T_{\rm eff}\sim 11\,400$ K (upper panel) and  for $0.1921 M_{\sun}$ 
pre-WD models at the same $T_{\rm eff}$ (lower panel). 
Different symbols connected with thin lines are associated with
models with different He abundance in the driving region, 
including that with canonical He abundance and those models 
(artificially modified) with increasing $X_{\rm He}$ values
in that region (see text). 
The large numerical range spanned by $\eta$ is appropriately
scaled for a better graphical representation. The vertical dashed lines 
correspond to the periods at $\sim 381$ s, $\sim 406$ s, and $\sim 421$ s 
observed in WASP J0247$-$25B.}
\label{figure_15} 
\end{center}
\end{figure}

For  WASP J0247$-$25B we have $T_{\rm eff}\sim 11\,380$ K and 
$\log g= 4.58$ \citep{2013Natur.498..463M}. 
As we can see from Fig. \ref{figure_01}, the evolutionary tracks 
passing close to the star are those 
with masses $M_{\star}= 0.1868 M_{\sun}$ and  $M_{\star}= 0.1921 M_{\sun}$.
In Fig. \ref{figure_15} we display the normalized growth rates 
$\eta$ for $\ell= 1$  modes as a function of the pulsation periods for  
$0.1868 M_{\sun}$ pre-WD models  at $T_{\rm eff}\sim 11\,400$ K 
(upper panel) and  for $0.1921 M_{\sun}$ pre-WD models at the 
same $T_{\rm eff}$ (lower panel). 
The vertical dashed lines correspond to the periods 
observed in WASP J0247$-$25B.
In the figure, we use different symbols (connected with thin lines) 
to show the resulting $\eta$ values for models with increasing  He abundance 
at the driving region. 
The ``canonical'' He abundance (that predicted by the evolutionary 
computations) is $X_{\rm He}= 0.536$, being the corresponding 
H abundance of $X_{\rm H}= 0.453$. As can be seen, in this case
(orange triangles up) all the pulsation modes are stable ($\eta < 0$).
The same is true when we increase $X_{\rm He}$ to 0.60. For 
$X_{\rm H}= 0.65$ we obtain some unstable modes ($\eta > 0$), 
and the number of unstable modes increases as we increase 
$X_{\rm He}$. Similar results are depicted in the lower panel,
corresponding to models with $M_{\star}= 0.1921 M_{\sun}$.
From the figure  it is apparent that, in the case 
of the $0.1868 M_{\sun}$ models (upper panel), in order to 
replicate the periods exhibited by WASP J0247$-$25B, the   
He abundance at the driving zone must be $X_{\rm He} \gtrsim 0.70$
($X_{\rm H}\lesssim 0.29$). Note that the range of excited periods
predicted by our computations when $X_{\rm He}= 0.80$ 
($370 \lesssim \Pi \lesssim 1150$ s) is much  wider than the range of 
periods shown 
by the star ($380\lesssim  \Pi \lesssim 420$ s). For this case, 
the three  unstable modes with periods close 
to the observed ones are $\ell= 1$ $p$ mixed modes 
with $k= 5$ ($\Pi\sim 431$ s), 
$k= 6$ ($\Pi \sim 398$ s), and $k= 7$ ($\Pi \sim 372$ s)\footnote{
We have not made any attempt to fit exactly the observed periods by 
finely tunning the abundances of He in the driving region, nor 
considering other harmonic degrees apart from $\ell= 1$.}.  
On the other hand, in the case of the $0.1921 M_{\sun}$ models 
(lower panel), it seems that there is no way to get a range of 
excited periods compatible with that exhibited by WASP J0247$-$25B. 
This is because the excited modes in our models have 
periods longer than the periods of the star, 
irrespective of the value that we adopt for $X_{\rm He}$ at the driving 
region.

From the above analysis, we conclude that the driving 
region ($\log T\sim 4.7$) of the star WASP J0247$-$25B must be 
characterized by a  much higher He abundance than that 
predicted by our pre-WD evolutionary models. The 
He abundances of pre-WD stars can depend on details of the previous
binary evolution, such as different donor star masses of the initial 
binary system and/or the beginning of mass loss at different times 
than considered in our computations. The exploration of these issues 
is beyond the scope of this paper.

\subsection{KIC 9164561}

Following the reverse approach than for WASP J0247$-$25B, 
here we explore how low the He abundance at the 
driving region must be as to prevent pulsational instabilities in the 
nonvariable star KIC 9164561. 
For a $0.1706 M_{\sun}$ pre-WD model  at $T_{\rm eff}\sim 10\,400$ K,
which is very close to the location of this star 
in the $T_{\rm eff} - \log g$ plane
(see Fig. \ref{figure_01}), we found that KIC 9164561 must have 
$X_{\rm He} \lesssim 0.46$ in order to avoid pulsational excitation.
This is $\sim 25 \%$ lower than the canonical value predicted 
by the previous evolution ($X_{\rm He}= 0.584$) for the 
$0.1706 M_{\sun}$ sequence.

\section{Summary and conclusions}
\label{conclusions}

In this paper, we have presented a detailed linear pulsation stability
analysis of pre-WD stars employing the set of 
state-of-the-art evolutionary models of \citet{2013A&A...557A..19A}.
This is the third paper in a series on pulsating low-mass, He-core 
WDs (including ELM WDs),  with the first one focused
on the adiabatic properties \citep{2014A&A...569A.106C}
 and the second one devoted to the nonadiabatic pulsation 
stability features of these stars \citep{2016A&A...585A...1C}. 
In the present paper, we explore the stability pulsation 
properties of low-mass He-core stellar models at phases prior 
to the WD evolution, that 
is, from effective temperatures $\sim 6000$ K until 
the stages before the stars reach their maximum effective temperature
at the beginning of the first cooling branch. This study was inspired by 
the discovery of the first two  pulsating  stars that could be
the precursors of pulsating  low-mass (including ELM)  WDs
WASP J0247$-$25B and WASP J1628$+$10B 
\citep{2013Natur.498..463M,2014MNRAS.444..208M}. 
We assess the pulsational stability of radial ($\ell= 0$) and 
nonradial ($\ell= 1, 2$) $g$ and $p$ modes for stellar models belonging 
to a  set of 11 evolutionary sequences extracted from the computations of 
 \citet{2013A&A...557A..19A}, with masses in the range 
$0.1554-0.2724M_{\sun}$. 

Our main findings are summarized below:

\begin{itemize}

\item[-] For stellar models in which element diffusion is 
neglected, we have confirmed and explored in detail 
a new instability strip in the
domain of low gravities and low effective temperatures of 
the $T_{\rm eff}-\log g$ diagram, where low-mass pre-WD 
stars ---the precursors of low-mass and ELM WDs--- are currently 
found. The destabilized modes are radial and nonradial $p$, $g$, and 
$p$-$g$ mixed modes (see Figs. \ref{figure_02}, \ref{figure_03}, 
and \ref{figure_04}), excited by the $\kappa-\gamma$ mechanism acting mainly
at the He$^+-$He$^{++}$  partial ionization region 
($\log T \sim 4.7$), with 
non-negligible contributions also from the He$-$He$^{+}$ ($\log T \sim 4.42$), 
and H$-$H$^+$ ($\log T \sim 4.15$) partial ionization zones at low effective 
temperatures ($T_{\rm eff} \lesssim 8000-8500$ K, 
see Fig. \ref{figure_05}). \citet{2013MNRAS.435..885J} have found
pulsation instability of radial modes due to the same driving 
mechanism, using stellar models completely independent of those employed
in this work. 

\item[-] The nonradial $\ell= 1$ and $\ell= 2$ blue edges of the instability 
strip, which are almost identical with each other, are hotter 
and somewhat less steep than the radial ($\ell= 0$) blue 
edge (see Fig. \ref{figure_01}). 

\item[-] Regarding the ranges of periods of the excited modes, 
they are strongly dependent on the value of the stellar mass
(Figs. \ref{figure_08} and \ref{figure_09}). For nonradial dipole
modes, the longest 
excited periods reach values in excess of $\sim 15\,000$ s 
for the highest-mass sequence showing unstable modes 
($M_{\star}= 0.2026 M_{\sun}$), and
this limit strongly decreases to $\sim 3000$ s for the less massive
sequence ($M_{\star}= 0.1554 M_{\sun})$ considered. 
In turn, the shortest excited periods range from 
$\sim 50$ s for $M_{\star}= 0.1554 M_{\sun}$, to
$\sim 3000$ s for $M_{\star}= 0.2026M_{\sun}$. In the case of radial 
modes, much fewer modes became unstable, with periods in the
range $\sim 70-2000$ s for models with $M_{\star}= 0.1554 M_{\sun}$,
and $\gtrsim 10\,000$ s for models with $M_{\star}= 0.1554 M_{\sun}$.
For high $T_{\rm eff}$, 
radial modes with $k = 0, 1, 2, 3, 4$ are excited for the sequence
with $M_{\star}= 0.1554 M_{\sun}$, whereas for sequences more massive
than $0.1822 M_{\sun}$, only the fundamental radial ($k= 0$) mode is
excited. For lower $T_{\rm eff}$, a modestly larger number of radial 
modes  are excited.

\item[-] As for the strength of the pulsational 
instabilities, generally the most excited modes 
(that is, with the shortest e-folding times) are those
$g$ and $p$ modes experiencing avoided crossings, characterized 
by short periods ($\tau_{\rm e}\sim 0.01-1$ yr), as can be seen 
in Fig. \ref{figure_07}. In contrast, modes with
longer periods are much more weakly destabilized, involving
longer e-folding times ($\tau_{\rm e}\sim 10-10^6$ yr), but still 
much shorter than the evolutionary timescales.

\item[-] A direct comparison of our findings 
with the results of  \citet{2013MNRAS.435..885J} is not possible
because both stability analyses are based on very different
stellar models and distinct pulsation codes. We can, however,
compare our instability domain (without diffusion, with 
$0.49 \lesssim X_{\rm He} \lesssim 0.62$) with 
the set of nonadiabatic models of \citet{2013MNRAS.435..885J}
in the case in which their envelope models have $X_{\rm He}= 
0.48$ ($X_{\rm H}= 0.50$) (see Fig. \ref{figure_10}). We 
found satisfactory agreement, although admittedly, 
the radial-mode blue edge of \citet{2013MNRAS.435..885J} 
is significantly hotter than ours. 
In addition, we found far fewer excited radial modes than
did \citet{2013MNRAS.435..885J}. 

\item[-] When time-dependent element diffusion is taken into account
in our models, the instability domain becomes substantially 
narrower and the blue edge is much cooler than in the case in which 
diffusion is neglected (see Fig. \ref{figure_11}). In this case,  
by virtue of the 
driving region at $\log T \sim 4.7$ becoming rapidly depleted in 
He, the modes are excited 
mainly by the $\kappa-\gamma$ mechanism acting at the 
H$-$H$^+$ partial ionization zone. In view of the narrowness 
of the effective temperature range in which pulsational 
instability is found, and because none of the two pre-ELMV 
stars known to date are
accounted for by these computations, we conclude that
element diffusion does not occur at these evolutionary stages
if the $\kappa-\gamma$ mechanism is responsible for the 
excitation of the observed pulsations.
Processes that might prevent diffusion could include 
stellar rotation and/or mass loss. This is a very important issue
that deserves to be explored in depth.

\item[-] Our computations are successful in explaining the location  of
  the pre-ELMV WD star WASP J1628$+$10B in the $T_{\rm eff}-\log g$
  plane (Fig. \ref{figure_01}) and also account for the  observed
  range of periods exhibited by this star  (Fig. \ref{figure_14}).  On
  the other hand, our nonadiabatic models fail to explain  the
  existence of the pre-ELMV WD star WASP J0247$-$25B. 
  This could indicate  that the He abundance 
  in the driving   region of this star  must be much larger than 
  our evolutionary computations predict, and that the mass of  
  the donor star in the initial binary system could be different 
  from that considered in our computations ($1 M_{\sun}$). 
  Alternatively, the pulsations of this star could be due to a 
  different excitation mechanism than the $\kappa-\gamma$ mechanism 
  considered here.
  In order to explore what the actual He
  abundance should be to get pulsational instability through the 
  $\kappa-\gamma$ mechanism at the 
  right $T_{\rm eff}$ and   $\log g$, with the range of periods displayed by this
  star, we artificially modify $X_{\rm He}$ at the region of the
  opacity bump due to the He${^+}-$He$^{++}$ partial ionization zone.
  We found that the star should have $X_{\rm He} \gtrsim 0.70$
  ($X_{\rm H}\lesssim 0.29$) if its effective temperature and  gravity
  are $T_{\rm eff}= 11\,380$ K and $\log g= 4.58$ (Figs. \ref{figure_15}).  
  This is in good agreement with the results of \citet{2013MNRAS.435..885J}.

\item[-] The nonvariable pre-ELM WDs considered fall outside our 
theoretical instability domain, except KIC 916456, which 
is located well within the predicetd 
nonradial ($\ell= 1, 2$) instability region 
(see Fig. \ref{figure_01}). We found 
that, in order to prevent pulsational excitation through the 
$\kappa-\gamma$ mechanism, the He abundance at the driving zone of this star 
must be $X_{\rm He} \lesssim 0.46$.

\end{itemize}

The discovery of two pulsating pre-ELMV WDs reported by 
\citet{2013Natur.498..463M,2014MNRAS.444..208M} indicates 
the possible existence of a new class of pulsating 
stars in the late stages of evolution of low-mass WDs. 
The full exploration of the nonadiabatic properties of stellar models through
this domain reported in the present paper, along with the study by 
\citet{2013MNRAS.435..885J}, constitute the first solid basis
to interpret present and future observations of pulsating pre-ELMV WDs.
New discoveries of additional members of this new class 
of pulsating stars and their analysis in the context of the present 
theoretical background  will allow to shed light on the 
evolutionary history of their progenitor stars.

\begin{acknowledgements}
We wish to thank our anonymous referee for the constructive
comments and suggestions that greatly improved the original version of
the paper. Part of this work was supported by AGENCIA through the Programa de
Modernizaci\'on Tecnol\'ogica BID 1728/OC-AR, and by the PIP
112-200801-00940 grant from CONICET. AMS is partially supported
by grants ESP2014-56003-R (MINECO) and 2014SGR-1458 (Generalitat of
Catalunya). The Armagh Observatory is funded by direct grant 
from the Northern Ireland Department of Culture Arts and Leisure. 
This research has made use of NASA Astrophysics Data System.    
\end{acknowledgements}


\bibliographystyle{aa} 
\bibliography{paper-aph} 

\begin{thebibliography}{71}
\expandafter\ifx\csname natexlab\endcsname\relax\def\natexlab#1{#1}\fi

\bibitem[{{Aizenman} {et~al.}(1977){Aizenman}, {Smeyers}, \&
  {Weigert}}]{1977A&A....58...41A}
{Aizenman}, M., {Smeyers}, P., \& {Weigert}, A. 1977, \aap, 58, 41

\bibitem[{{Althaus} \& {Benvenuto}(2000)}]{2000MNRAS.317..952A}
{Althaus}, L.~G. \& {Benvenuto}, O.~G. 2000, \mnras, 317, 952

\bibitem[{{Althaus} {et~al.}(2010){Althaus}, {C{\'o}rsico}, {Isern}, \&
  {Garc{\'{\i}}a-Berro}}]{review}
{Althaus}, L.~G., {C{\'o}rsico}, A.~H., {Isern}, J., \& {Garc{\'{\i}}a-Berro},
  E. 2010, \aapr, 18, 471

\bibitem[{{Althaus} {et~al.}(2013){Althaus}, {Miller Bertolami}, \&
  {C{\'o}rsico}}]{2013A&A...557A..19A}
{Althaus}, L.~G., {Miller Bertolami}, M.~M., \& {C{\'o}rsico}, A.~H. 2013,
  \aap, 557, A19

\bibitem[{{Althaus} {et~al.}(2009){Althaus}, {Panei}, {Romero}, {Rohrmann},
  {C{\'o}rsico}, {Garc{\'{\i}}a-Berro}, \& {Miller
  Bertolami}}]{2009A&A...502..207A}
{Althaus}, L.~G., {Panei}, J.~A., {Romero}, A.~D., {et~al.} 2009, \aap, 502,
  207

\bibitem[{{Althaus} {et~al.}(2001){Althaus}, {Serenelli}, \&
  {Benvenuto}}]{2001ApJ...554.1110A}
{Althaus}, L.~G., {Serenelli}, A.~M., \& {Benvenuto}, O.~G. 2001, \apj, 554,
  1110

\bibitem[{{Althaus} {et~al.}(2005){Althaus}, {Serenelli}, {Panei},
  {C{\'o}rsico}, {Garc{\'{\i}}a-Berro}, \&
  {Sc{\'o}ccola}}]{2005A&A...435..631A}
{Althaus}, L.~G., {Serenelli}, A.~M., {Panei}, J.~A., {et~al.} 2005, \aap, 435,
  631

\bibitem[{{Antoci} {et~al.}(2014){Antoci}, {Cunha}, {Houdek}, {Kjeldsen},
  {Trampedach}, {Handler}, {L{\"u}ftinger}, {Arentoft}, \&
  {Murphy}}]{2014ApJ...796..118A}
{Antoci}, V., {Cunha}, M., {Houdek}, G., {et~al.} 2014, \apj, 796, 118

\bibitem[{{Antoci} {et~al.}(2011){Antoci}, {Handler}, {Campante}, {Thygesen},
  {Moya}, {Kallinger}, {Stello}, {Grigahc{\`e}ne}, {Kjeldsen}, {Bedding},
  {L{\"u}ftinger}, {Christensen-Dalsgaard}, {Catanzaro}, {Frasca}, {De Cat},
  {Uytterhoeven}, {Bruntt}, {Houdek}, {Kurtz}, {Lenz}, {Kaiser}, {van Cleve},
  {Allen}, \& {Clarke}}]{2011Natur.477..570A}
{Antoci}, V., {Handler}, G., {Campante}, T.~L., {et~al.} 2011, \nat, 477, 570

\bibitem[{{Bell} {et~al.}(2015){Bell}, {Kepler}, {Montgomery}, {Hermes},
  {Harrold}, \& {Winget}}]{2015ASPC..493..217B}
{Bell}, K.~J., {Kepler}, S.~O., {Montgomery}, M.~H., {et~al.} 2015, in
  Astronomical Society of the Pacific Conference Series, Vol. 493, 19th
  European Workshop on White Dwarfs, ed. P.~{Dufour}, P.~{Bergeron}, \&
  G.~{Fontaine}, 217

\bibitem[{{Bohm} \& {Cassinelli}(1971)}]{1971A&A....12...21B}
{Bohm}, K.~H. \& {Cassinelli}, J. 1971, \aap, 12, 21

\bibitem[{{B{\"o}hm-Vitense}(1958)}]{1958ZA.....46..108B}
{B{\"o}hm-Vitense}, E. 1958, \zap, 46, 108

\bibitem[{{Breton} {et~al.}(2012){Breton}, {Rappaport}, {van Kerkwijk}, \&
  {Carter}}]{2012ApJ...748..115B}
{Breton}, R.~P., {Rappaport}, S.~A., {van Kerkwijk}, M.~H., \& {Carter}, J.~A.
  2012, \apj, 748, 115

\bibitem[{{Brickhill}(1991)}]{1991MNRAS.251..673B}
{Brickhill}, A.~J. 1991, \mnras, 251, 673

\bibitem[{{Brown} {et~al.}(2013){Brown}, {Kilic}, {Allende Prieto},
  {Gianninas}, \& {Kenyon}}]{2013ApJ...769...66B}
{Brown}, W.~R., {Kilic}, M., {Allende Prieto}, C., {Gianninas}, A., \&
  {Kenyon}, S.~J. 2013, \apj, 769, 66

\bibitem[{{Brown} {et~al.}(2010){Brown}, {Kilic}, {Allende Prieto}, \&
  {Kenyon}}]{2010ApJ...723.1072B}
{Brown}, W.~R., {Kilic}, M., {Allende Prieto}, C., \& {Kenyon}, S.~J. 2010,
  \apj, 723, 1072

\bibitem[{{Brown} {et~al.}(2012){Brown}, {Kilic}, {Allende Prieto}, \&
  {Kenyon}}]{2012ApJ...744..142B}
---. 2012, \apj, 744, 142

\bibitem[{{Burgers}(1969)}]{1969fecg.book.....B}
{Burgers}, J.~M. 1969, {Flow Equations for Composite Gases} (New York: Academic
  Press)

\bibitem[{{Carter} {et~al.}(2011){Carter}, {Rappaport}, \&
  {Fabrycky}}]{2011ApJ...728..139C}
{Carter}, J.~A., {Rappaport}, S., \& {Fabrycky}, D. 2011, \apj, 728, 139

\bibitem[{{Cassisi} {et~al.}(2007){Cassisi}, {Potekhin}, {Pietrinferni},
  {Catelan}, \& {Salaris}}]{2007ApJ...661.1094C}
{Cassisi}, S., {Potekhin}, A.~Y., {Pietrinferni}, A., {Catelan}, M., \&
  {Salaris}, M. 2007, \apj, 661, 1094

\bibitem[{{Christensen-Dalsgaard} \& {Houdek}(2010)}]{2010Ap&SS.328...51C}
{Christensen-Dalsgaard}, J. \& {Houdek}, G. 2010, \apss, 328, 51

\bibitem[{{C{\'o}rsico} \& {Althaus}(2006)}]{2006A&A...454..863C}
{C{\'o}rsico}, A.~H. \& {Althaus}, L.~G. 2006, \aap, 454, 863

\bibitem[{{C{\'o}rsico} \& {Althaus}(2014{\natexlab{a}})}]{2014A&A...569A.106C}
---. 2014{\natexlab{a}}, \aap, 569, A106

\bibitem[{{C{\'o}rsico} \& {Althaus}(2014{\natexlab{b}})}]{2014ApJ...793L..17C}
---. 2014{\natexlab{b}}, \apjl, 793, L17

\bibitem[{{C{\'o}rsico} \& {Althaus}(2016)}]{2016A&A...585A...1C}
---. 2016, \aap, 585, A1

\bibitem[{{C{\'o}rsico} {et~al.}(2006){C{\'o}rsico}, {Althaus}, \& {Miller
  Bertolami}}]{2006A&A...458..259C}
{C{\'o}rsico}, A.~H., {Althaus}, L.~G., \& {Miller Bertolami}, M.~M. 2006,
  \aap, 458, 259

\bibitem[{{C{\'o}rsico} {et~al.}(2012){C{\'o}rsico}, {Romero}, {Althaus}, \&
  {Hermes}}]{2012A&A...547A..96C}
{C{\'o}rsico}, A.~H., {Romero}, A.~D., {Althaus}, L.~G., \& {Hermes}, J.~J.
  2012, \aap, 547, A96

\bibitem[{{Cox}(1968)}]{1968pss..book.....C}
{Cox}, J.~P. 1968, {Principles of stellar structure - Vol.1: Physical
  principles; Vol.2: Applications to stars}

\bibitem[{{Cox}(1980)}]{1980tsp..book.....C}
---. 1980, {Theory of stellar pulsation}

\bibitem[{{Deheuvels} \& {Michel}(2010)}]{2010Ap&SS.328..259D}
{Deheuvels}, S. \& {Michel}, E. 2010, \apss, 328, 259

\bibitem[{{Driebe} {et~al.}(1998){Driebe}, {Schoenberner}, {Bloecker}, \&
  {Herwig}}]{1998A&A...339..123D}
{Driebe}, T., {Schoenberner}, D., {Bloecker}, T., \& {Herwig}, F. 1998, \aap,
  339, 123

\bibitem[{{Fontaine} \& {Brassard}(2008)}]{2008PASP..120.1043F}
{Fontaine}, G. \& {Brassard}, P. 2008, \pasp, 120, 1043

\bibitem[{{Gianninas} {et~al.}(2014){Gianninas}, {Dufour}, {Kilic}, {Brown},
  {Bergeron}, \& {Hermes}}]{2014ApJ...794...35G}
{Gianninas}, A., {Dufour}, P., {Kilic}, M., {et~al.} 2014, \apj, 794, 35

\bibitem[{{Gianninas} {et~al.}(2015){Gianninas}, {Kilic}, {Brown}, {Canton}, \&
  {Kenyon}}]{2015ApJ...812..167G}
{Gianninas}, A., {Kilic}, M., {Brown}, W.~R., {Canton}, P., \& {Kenyon}, S.~J.
  2015, \apj, 812, 167

\bibitem[{{Haft} {et~al.}(1994){Haft}, {Raffelt}, \&
  {Weiss}}]{1994ApJ...425..222H}
{Haft}, M., {Raffelt}, G., \& {Weiss}, A. 1994, \apj, 425, 222

\bibitem[{{Hermes} {et~al.}(2013{\natexlab{a}}){Hermes}, {Montgomery},
  {Gianninas}, {Winget}, {Brown}, {Harrold}, {Bell}, {Kenyon}, {Kilic}, \&
  {Castanheira}}]{2013MNRAS.436.3573H}
{Hermes}, J.~J., {Montgomery}, M.~H., {Gianninas}, A., {et~al.}
  2013{\natexlab{a}}, \mnras, 436, 3573

\bibitem[{{Hermes} {et~al.}(2013{\natexlab{b}}){Hermes}, {Montgomery},
  {Winget}, {Brown}, {Gianninas}, {Kilic}, {Kenyon}, {Bell}, \&
  {Harrold}}]{2013ApJ...765..102H}
{Hermes}, J.~J., {Montgomery}, M.~H., {Winget}, D.~E., {et~al.}
  2013{\natexlab{b}}, \apj, 765, 102

\bibitem[{{Hermes} {et~al.}(2012){Hermes}, {Montgomery}, {Winget}, {Brown},
  {Kilic}, \& {Kenyon}}]{2012ApJ...750L..28H}
---. 2012, \apjl, 750, L28

\bibitem[{{Iglesias} \& {Rogers}(1996)}]{1996ApJ...464..943I}
{Iglesias}, C.~A. \& {Rogers}, F.~J. 1996, \apj, 464, 943

\bibitem[{{Istrate} {et~al.}(2014){Istrate}, {Tauris}, \&
  {Langer}}]{2014A&A...571A..45I}
{Istrate}, A.~G., {Tauris}, T.~M., \& {Langer}, N. 2014, \aap, 571, A45

\bibitem[{{Itoh} {et~al.}(1996){Itoh}, {Hayashi}, {Nishikawa}, \&
  {Kohyama}}]{1996ApJS..102..411I}
{Itoh}, N., {Hayashi}, H., {Nishikawa}, A., \& {Kohyama}, Y. 1996, \apjs, 102,
  411

\bibitem[{{Jeffery} \& {Saio}(2013)}]{2013MNRAS.435..885J}
{Jeffery}, C.~S. \& {Saio}, H. 2013, \mnras, 435, 885

\bibitem[{{Kawaler}(1993)}]{1993ApJ...404..294K}
{Kawaler}, S.~D. 1993, \apj, 404, 294

\bibitem[{{Kepler} {et~al.}(2007){Kepler}, {Kleinman}, {Nitta}, {Koester},
  {Castanheira}, {Giovannini}, {Costa}, \& {Althaus}}]{2007MNRAS.375.1315K}
{Kepler}, S.~O., {Kleinman}, S.~J., {Nitta}, A., {et~al.} 2007, \mnras, 375,
  1315

\bibitem[{{Kepler} {et~al.}(2015){Kepler}, {Pelisoli}, {Koester}, {Ourique},
  {Kleinman}, {Romero}, {Nitta}, {Eisenstein}, {Costa}, {K{\"u}lebi}, {Jordan},
  {Dufour}, {Giommi}, \& {Rebassa-Mansergas}}]{2015MNRAS.446.4078K}
{Kepler}, S.~O., {Pelisoli}, I., {Koester}, D., {et~al.} 2015, \mnras, 446,
  4078

\bibitem[{{Kilic} {et~al.}(2011){Kilic}, {Brown}, {Allende Prieto},
  {Ag{\"u}eros}, {Heinke}, \& {Kenyon}}]{2011ApJ...727....3K}
{Kilic}, M., {Brown}, W.~R., {Allende Prieto}, C., {et~al.} 2011, \apj, 727, 3

\bibitem[{{Kilic} {et~al.}(2012){Kilic}, {Brown}, {Allende Prieto}, {Kenyon},
  {Heinke}, {Ag{\"u}eros}, \& {Kleinman}}]{2012ApJ...751..141K}
---. 2012, \apj, 751, 141

\bibitem[{{Kilic} {et~al.}(2015){Kilic}, {Hermes}, {Gianninas}, \&
  {Brown}}]{2015MNRAS.446L..26K}
{Kilic}, M., {Hermes}, J.~J., {Gianninas}, A., \& {Brown}, W.~R. 2015, \mnras,
  446, L26

\bibitem[{{Kleinman} {et~al.}(2013){Kleinman}, {Kepler}, {Koester}, {Pelisoli},
  {Pe{\c c}anha}, {Nitta}, {Costa}, {Krzesinski}, {Dufour}, {Lachapelle},
  {Bergeron}, {Yip}, {Harris}, {Eisenstein}, {Althaus}, \&
  {C{\'o}rsico}}]{2013ApJS..204....5K}
{Kleinman}, S.~J., {Kepler}, S.~O., {Koester}, D., {et~al.} 2013, \apjs, 204, 5

\bibitem[{{Koester} {et~al.}(2009){Koester}, {Voss}, {Napiwotzki},
  {Christlieb}, {Homeier}, {Lisker}, {Reimers}, \&
  {Heber}}]{2009A&A...505..441K}
{Koester}, D., {Voss}, B., {Napiwotzki}, R., {et~al.} 2009, \aap, 505, 441

\bibitem[{{Lee} \& {Bradley}(1993)}]{1993ApJ...418..855L}
{Lee}, U. \& {Bradley}, P.~A. 1993, \apj, 418, 855

\bibitem[{{Magni} \& {Mazzitelli}(1979)}]{1979A&A....72..134M}
{Magni}, G. \& {Mazzitelli}, I. 1979, \aap, 72, 134

\bibitem[{{Maxted} {et~al.}(2011){Maxted}, {Anderson}, {Burleigh}, {Collier
  Cameron}, {Heber}, {G{\"a}nsicke}, {Geier}, {Kupfer}, {Marsh}, {Nelemans},
  {O'Toole}, {{\O}stensen}, {Smalley}, \& {West}}]{2011MNRAS.418.1156M}
{Maxted}, P.~F.~L., {Anderson}, D.~R., {Burleigh}, M.~R., {et~al.} 2011,
  \mnras, 418, 1156

\bibitem[{{Maxted} {et~al.}(2014){Maxted}, {Serenelli}, {Marsh}, {Catal{\'a}n},
  {Mahtani}, \& {Dhillon}}]{2014MNRAS.444..208M}
{Maxted}, P.~F.~L., {Serenelli}, A.~M., {Marsh}, T.~R., {et~al.} 2014, \mnras,
  444, 208

\bibitem[{{Maxted} {et~al.}(2013){Maxted}, {Serenelli}, {Miglio}, {Marsh},
  {Heber}, {Dhillon}, {Littlefair}, {Copperwheat}, {Smalley}, {Breedt}, \&
  {Schaffenroth}}]{2013Natur.498..463M}
{Maxted}, P.~F.~L., {Serenelli}, A.~M., {Miglio}, A., {et~al.} 2013, \nat, 498,
  463

\bibitem[{{Osaki}(1975)}]{1975PASJ...27..237O}
{Osaki}, J. 1975, \pasj, 27, 237

\bibitem[{{Panei} {et~al.}(2007){Panei}, {Althaus}, {Chen}, \&
  {Han}}]{2007MNRAS.382..779P}
{Panei}, J.~A., {Althaus}, L.~G., {Chen}, X., \& {Han}, Z. 2007, \mnras, 382,
  779

\bibitem[{{Rappaport} {et~al.}(2015){Rappaport}, {Nelson}, {Levine},
  {Sanchis-Ojeda}, {Gandolfi}, {Nowak}, {Palle}, \&
  {Prsa}}]{2015ApJ...803...82R}
{Rappaport}, S., {Nelson}, L., {Levine}, A., {et~al.} 2015, \apj, 803, 82

\bibitem[{{Saio}(2013)}]{2013EPJWC..4305005S}
{Saio}, H. 2013, in European Physical Journal Web of Conferences, Vol.~43,
  European Physical Journal Web of Conferences, 5005

\bibitem[{{Saio} {et~al.}(1983){Saio}, {Winget}, \&
  {Robinson}}]{1983ApJ...265..982S}
{Saio}, H., {Winget}, D.~E., \& {Robinson}, E.~L. 1983, \apj, 265, 982

\bibitem[{{Sarna} {et~al.}(2000){Sarna}, {Ergma}, \& {Ger{\v s}kevit{\v
  s}-Antipova}}]{2000MNRAS.316...84S}
{Sarna}, M.~J., {Ergma}, E., \& {Ger{\v s}kevit{\v s}-Antipova}, J. 2000,
  \mnras, 316, 84

\bibitem[{{Scuflaire}(1974)}]{1974A&A....36..107S}
{Scuflaire}, R. 1974, \aap, 36, 107

\bibitem[{{Steinfadt} {et~al.}(2010){Steinfadt}, {Bildsten}, \&
  {Arras}}]{2010ApJ...718..441S}
{Steinfadt}, J.~D.~R., {Bildsten}, L., \& {Arras}, P. 2010, \apj, 718, 441

\bibitem[{{Strittmatter} \& {Wickramasinghe}(1971)}]{1971MNRAS.152...47S}
{Strittmatter}, P.~A. \& {Wickramasinghe}, D.~T. 1971, \mnras, 152, 47

\bibitem[{{Tassoul} {et~al.}(1990){Tassoul}, {Fontaine}, \&
  {Winget}}]{1990ApJS...72..335T}
{Tassoul}, M., {Fontaine}, G., \& {Winget}, D.~E. 1990, \apjs, 72, 335

\bibitem[{{Tremblay} {et~al.}(2011){Tremblay}, {Bergeron}, \&
  {Gianninas}}]{2011ApJ...730..128T}
{Tremblay}, P.-E., {Bergeron}, P., \& {Gianninas}, A. 2011, \apj, 730, 128

\bibitem[{{Unglaub} \& {Bues}(2000)}]{2000A&A...359.1042U}
{Unglaub}, K. \& {Bues}, I. 2000, \aap, 359, 1042

\bibitem[{{Unno} {et~al.}(1989){Unno}, {Osaki}, {Ando}, {Saio}, \&
  {Shibahashi}}]{1989nos..book.....U}
{Unno}, W., {Osaki}, Y., {Ando}, H., {Saio}, H., \& {Shibahashi}, H. 1989,
  {Nonradial oscillations of stars}, ed. T.~University~of Tokyo~Press

\bibitem[{{Van Grootel} {et~al.}(2013){Van Grootel}, {Fontaine}, {Brassard}, \&
  {Dupret}}]{2013ApJ...762...57V}
{Van Grootel}, V., {Fontaine}, G., {Brassard}, P., \& {Dupret}, M.-A. 2013,
  \apj, 762, 57

\bibitem[{{van Kerkwijk} {et~al.}(2010){van Kerkwijk}, {Rappaport}, {Breton},
  {Justham}, {Podsiadlowski}, \& {Han}}]{2010ApJ...715...51V}
{van Kerkwijk}, M.~H., {Rappaport}, S.~A., {Breton}, R.~P., {et~al.} 2010,
  \apj, 715, 51

\bibitem[{{Winget} \& {Kepler}(2008)}]{2008ARA&A..46..157W}
{Winget}, D.~E. \& {Kepler}, S.~O. 2008, \araa, 46, 157

\end{thebibliography}

\end{document}